\documentclass[preprint]{aastex63}

\usepackage{chemformula}
\let\ce\ch

\usepackage{comment}

\definecolor{darkorange}{RGB}{179,98,0}

\received{\today}
\revised{NaN}
\accepted{NaN}

\submitjournal{PSJ}

\shorttitle{The Clouds of Venus}
\shortauthors{Rimmer et al.}

\graphicspath{{./}{figures/}}

\begin{document}

\title{Hydroxide salts in the clouds of Venus: their effect on the sulfur cycle and cloud droplet pH}

\correspondingauthor{Paul B. Rimmer}
\email{pbr27@cam.ac.uk}

\author[0000-0002-7180-081X]{Paul B. Rimmer}
\affiliation{Department of Earth Sciences, University of Cambridge, Downing St, Cambridge CB2 3EQ, United Kingdom}
\affiliation{Cavendish Laboratory, University of Cambridge, JJ Thomson Ave, Cambridge CB3 0HE, United Kingdom}
\affiliation{MRC Laboratory of Molecular Biology, Francis Crick Ave, Cambridge CB2 0QH, United Kingdom}

\author{Sean Jordan}
\affiliation{Institute of Astronoomy, University of Cambridge, Madingley Rd, Cambridge CB3 0HA, United Kingdom}

\author{Tereza Constantinou}
\affiliation{Institute of Astronoomy, University of Cambridge, Madingley Rd, Cambridge CB3 0HA, United Kingdom}

\author{Peter Woitke}
\affiliation{SUPA, School of Physics \& Astronomy, University of St Andrews, St Andrews, KY16 9SS, UK}
\affiliation{Centre for Exoplanet Science, University of St Andrews, St Andrews, UK}

\author[0000-0002-8713-1446]{Oliver Shorttle}
\affiliation{Department of Earth Sciences, University of Cambridge, Downing St, Cambridge CB2 3EQ, United Kingdom}
\affiliation{Institute of Astronomy, University of Cambridge, Madingley Rd, Cambridge CB3 0HA, United Kingdom}

\author{Alessia Paschodimas}
\affiliation{Earth and Environmental Sciences, University of St Andrews, Irvine Building, North Street, St Andrews, KY16 9AL, United Kingdom}
\affiliation{Centre for Exoplanet Science, University of St Andrews, St Andrews, UK}

\author{Richard Hobbs}
\affiliation{Institute of Astronomy, University of Cambridge, Madingley Rd, Cambridge CB3 0HA, United Kingdom}


\begin{abstract}
The depletion of \ce{SO_2} and \ce{H_2O} in and above the clouds of Venus (45 -- 65 km) cannot be explained by known gas-phase chemistry and the observed composition of the atmosphere. We apply a full-atmosphere model of Venus to investigate three potential explanations for the \ce{SO_2} and \ce{H_2O} depletion: (1) varying the below-cloud water vapor (\ce{H_2O}), (2) varying the below-cloud sulfur dioxide (\ce{SO_2}), and (3) the incorporation of chemical reactions inside the sulfuric acid cloud droplets. We find that increasing the below-cloud \ce{H_2O} to explain the \ce{SO_2} depletion results in a cloud top that is 20 km too high, above-cloud \ce{O_2} three orders of magnitude greater than observational upper limits and no \ce{SO} above 80 km. The \ce{SO_2} depletion can be explained by decreasing the below-cloud \ce{SO_2} to $20\,{\rm ppm}$. The depletion of \ce{SO_2} in the clouds can also be explained by the \ce{SO_2} dissolving into the clouds, if the droplets contain hydroxide salts. These salts buffer the cloud pH. The amount of salts sufficient to explain the \ce{SO_2} depletion entail a droplet pH of $\sim 1$ at 50 km. Since sulfuric acid is constantly condensing out into the cloud droplets, there must be a continuous and pervasive flux of salts of $\approx 10^{-13} \, {\rm mol \, cm^{-2} \, s^{-1}}$ driving the cloud droplet chemistry. An atmospheric probe can test both of these explanations by measuring the pH of the cloud droplets and the concentrations of gas-phase \ce{SO_2} below the clouds.
\end{abstract}

\keywords{Venus --- Planetary atmospheres --- Sulfur cycle --- Sulfur dioxide --- Water vapor}

\section{Introduction} 
\label{sec:intro}

Both sulfur dioxide (\ce{SO_2}) and water vapor (\ce{H_2O}) are known to be depleted in the cloud layer of Venus \citep[see, e.g.,][]{Vandaele2017,Bierson2020}, and to vary in abundance above the cloud top by an order of magnitude or more both spatially \citep{Jessup2015,Encrenaz2019,Marcq2020}, and temporally in years-long cycles \citep{Marcq2013,Vandaele2017b}. Both of these species participate in Venus's atmospheric sulfur cycle \citep{Yung1982,Kras1982,Kras2007,Kras2010,Kras2012,Mills2007,Zhang2012,Bierson2020}. Their photo-destruction in the upper cloud layer (60 -- 70 km) leads to formation of sulfuric acid (\ce{H_2SO_4}), that condenses out and forms the clouds in Venus's atmosphere \citep{Yung1982}. The droplets rain out of the clouds at a height of $\lesssim 48$ km, where they evaporate \citep{Yung1982,Kras2007}. The \ce{H_2SO_4} then dissociates and replenishes \ce{SO_2} and \ce{H_2O} in the lower atmosphere \citep[e.g.,][]{Kras2007}. The behavior of all other known chemically reactive species in Venus's atmosphere is influenced by this cycle \citep{Kras2007,Kras2010,Kras2012}, and many of these species participate in this cycle. The sulfur cycle in the atmosphere of Venus establishes a strong and persistent redox gradient through the atmosphere of Venus. Venus is more reduced above the clouds and more oxidized below the clouds.

Though this cycle is central to the atmospheric chemistry of Venus, it is not fully understood, and no self-consistent full atmospheric model of Venus yet accounts for this cycle.

\begin{figure}[ht!]
\plotone{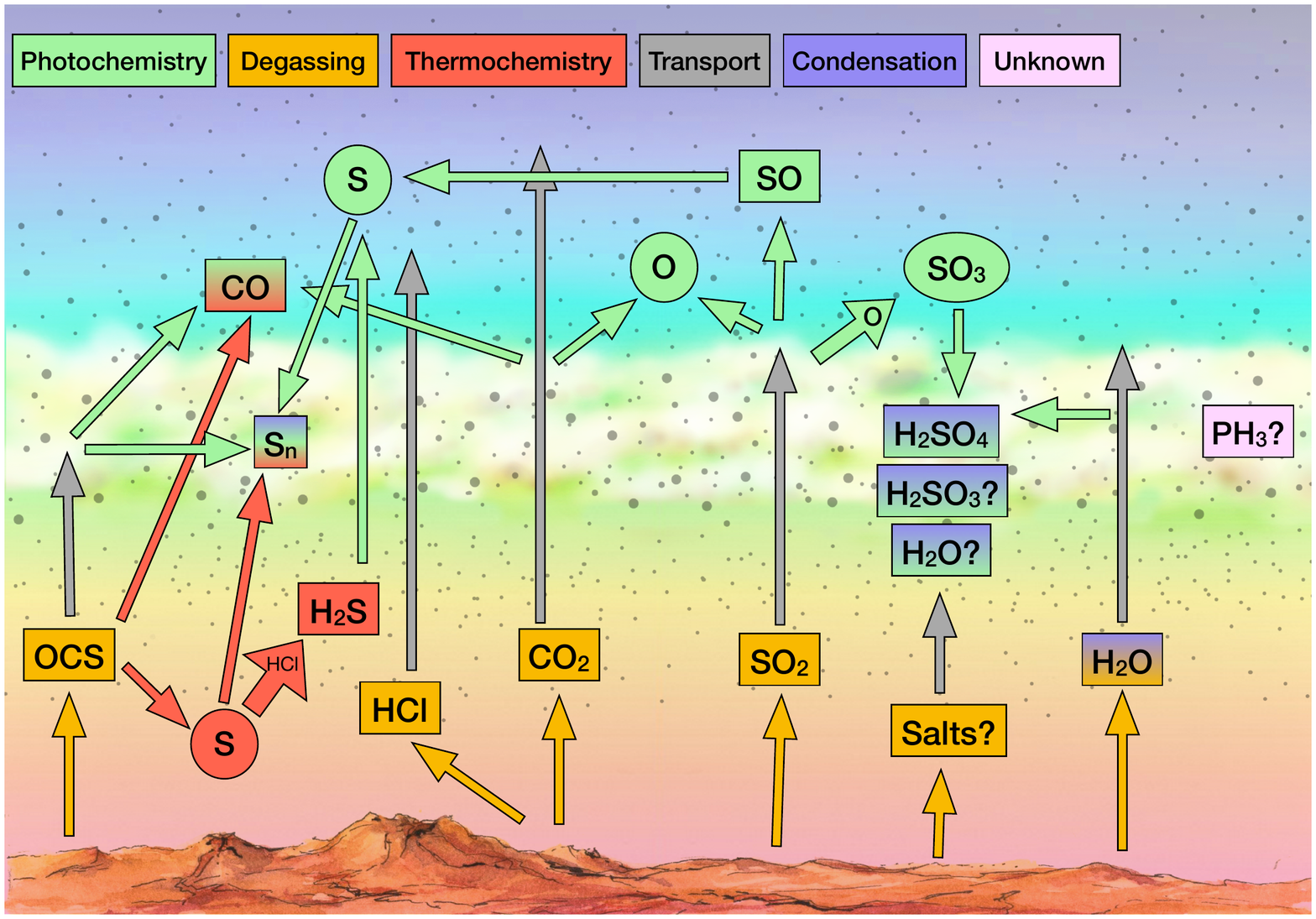}
\caption{A scheme of the sulfur cycle on Venus. We include a hypothetical mechanism for cloud buffering by volcanic release of salts into the clouds, or levitation of dust particles with salts by winds into the cloud layer. For simplicity, this figure only shows the upward transport of species relevant for formation of the cloud and initiation of droplet chemistry, and not the settling, rainout and evaporation of cloud droplets, needed to complete the sulfur cycle. \label{fig:cartoon}}
\end{figure}

There are several models of the lower atmosphere of Venus (0 -- 40 km) that account for the efficient evaporation of \ce{H_2SO_4} and the effect of its dissociation products, \ce{SO_3} and \ce{H_2O}, on the abundances of carbon monoxide (\ce{CO}), carbonyl sulfide (\ce{OCS}), and \ce{SO_2} \citep{Kras2007,Kras2013}. Other models describe the middle atmosphere of Venus (60 -- 120 km) \citep[e.g.,][]{Zhang2012}, investigating the chemistry above the clouds (60 -- 80 km) where \ce{SO_2} is depleted and then re-appears between 85 and 105 km \citep{Sandor2010,Belyaev2012}. \citet{Zhang2010} propose that night-side evaporation of \ce{H_2SO4} at 85 -- 105 km, followed by rapid displacement to the day side by strong winds and subsequent photodissociation, can explain this behavior. The distribution of upper atmospheric aerosols at day and night side \citep{Parkinson2015}, could be applied as constraints for the proposed mechanism of \citet{Zhang2010}. \citet{Pinto2021} suggest an alternative hypothesis, involving \ce{(SO)_2} chemistry. There has recently been a model of the upper clouds layer of Venus exploring the \ce{SO_2} depletion from $1\,\mathrm{ppm}$ to $\sim 10\,\mathrm{ppb}$, and the correlation with \ce{H_2O} abundances in the clouds \citep{Shao2020}. A diagram of the sulfur cycle and its connection to other trace atmospheric species in the atmosphere of Venus is shown in Figure \ref{fig:cartoon}.

At least three atmospheric models of Venus also exist, but they either do not predict the observed \ce{SO_2} depletion \citep{Yung2009}, do not couple the \ce{SO_2} depletion to the sulfur cycle \citep{Greaves2020}, or do not consider the \ce{H_2O} and \ce{SO_2} depletion in concert. The best current full-atmospheric model that accounts for the \ce{SO_2} depletion, from \citet{Bierson2020}, reproduces the \ce{SO_2} reasonably well, though only by fixing the \ce{H_2O} profile and by inhibiting the vertical transport within the clouds, as suggested by \citet{Marcq2018} and \citet{Bierson2020}.

To explain the depletion of sulfur dioxide and water in concert, either some unknown chemistry must take place within the cloud layer, or observations of lower atmospheric \ce{SO_2} and/or \ce{H_2O} must be mistaken. We explore both of these possibilities in this paper.

In Section \ref{sec:puzzle} we show why the sulfur cycle cannot be explained without either decreasing the amount of sulfur in the lower atmosphere or increasing the amount of hydrogen in the clouds, either by increasing the water vapor in the clouds or by transporting the hydrogen into the clouds in a different form. We propose that hydrogen could be contained either in aerosols that are lifted by winds into the clouds from the surface, delivered exogenously, or contained within the clouds within some unknown chemical species. In Section \ref{sec:model}, we discuss our full atmospheric model for Venus. We then show the results of our model if the observational constraints on \ce{SO_2} and \ce{H_2O} in the lower atmosphere are wrong (Sections \ref{sec:results-h2o} and \ref{sec:results-so2}) or if we introduce cloud chemistry (Section \ref{sec:results-clouds}). In Section \ref{sec:results-clouds} we also predict the effect of this source of hydrogen on the cloud chemistry, chiefly on how it would act as a pH buffer in the clouds. We discuss the implications of our results in Section \ref{sec:discussion}, particularly about how rainout and replenishment of hydrogen is needed to sustain the \ce{SO_2} gradient. We also speculate about possible sources of hydrogen and their delivery into the clouds, and ways of reconciling the changing cloud chemistry with observations. Section \ref{sec:conclusion} contains our conclusions.

\section{The Puzzle of Sulfur Depletion} 
\label{sec:puzzle}

There is good observational evidence that the concentration of \ce{SO_2}, the dominant sulfur-bearing species in Venus's atmosphere, varies by several orders of magnitude between altitudes of 40 -- 80 km (see Appendix \ref{app:obs}. The concentration of \ce{SO_2} is above 100 ppm at 40 km and between 1 and 100 ppb at 80 km, and we call the decrease in \ce{SO_2} with height the ``depletion'' of \ce{SO_2}. The standard explanation for this depletion predicts the major presumed constituent of the clouds of Venus: \ce{H2SO4}. Near the top of the clouds:
\begin{align}
\ce{CO_2} + h\nu &\rightarrow \ce{CO} + \ce{O} \label{eqn:co2-pd}\\
\ce{SO_2} + h\nu &\rightarrow \ce{SO} + \ce{O} \label{eqn:so2-pd}\\
\ce{SO_2} + \ce{O} + \ce{M} &\rightarrow \ce{SO_3} + \ce{M} \label{eqn:so2-o}\\
\ce{SO_3} + 2\ce{H_2O} &\rightarrow \ce{H_2SO_4} + \ce{H_2O} \label{eqn:so3-h2o}
\end{align}
The \ce{H_2SO_4} condenses out to droplets that then drop down to $\sim 40$ km, where the droplets evaporate. These droplets are predicted to make up the cloud layer. This mechanism ends up reducing the upper atmosphere by replacing \ce{CO_2} with \ce{CO} and \ce{SO_2} with \ce{SO}. The excess oxygen is bound up in the sulfuric acid which has condensed out of the atmosphere.

An oxygen atom is needed to form \ce{SO_3}, and this \ce{O} must come from either \ce{CO_2} or \ce{SO_2} because they are by far the most abundant \ce{O}-containing molecules. The formation of one molecule of \ce{H_2SO_4} is the destruction of one molecule of \ce{H_2O} and at least one molecule of \ce{SO_2} (the sum of Reactions (\ref{eqn:co2-pd}) and (\ref{eqn:so2-o})), at most two molecules of \ce{SO_2} (the sum of Reactions (\ref{eqn:so2-pd}) and (\ref{eqn:so2-o})). Between one and two molecules of \ce{SO_2} is lost with every molecule of \ce{H_2O} to make a molecule of \ce{H_2SO_4} this way, and so this mechanism predicts that the below-cloud \ce{H_2O} and \ce{SO_2} concentrations be within a factor of two of each other. Therefore, the maximum depletion of \ce{SO_2} in the atmosphere by this mechanism is equal to $\big(\chi + 1\big)\ce{[H_2O]}$, where \ce{[H_2O]} (cm$^{-3}$) is the atmospheric mixing ratio of \ce{H_2O} and $\chi$ is the fraction of \ce{H2SO4}-bound \ce{O} that was produced by \ce{SO_2} dissociation.

 The observational constraints, however, are $[\ce{SO_2}] \approx 150$ ppm and $[\ce{H_2O}] \approx 30$ ppm. Even if $\chi = 1$, and all water was converted to \ce{H_2SO_4}, the \ce{SO_2} would only be depleted by $\sim 20\%$. This is insufficient to explain the several-orders-of-magnitude depletion of \ce{SO_2}.

Therefore the \ce{SO_2} depletion is a puzzle for which there is no successful solution in the literature consistent with observations. This implies that either the observational constraints on \ce{H_2O} and/or \ce{SO_2} in the lower atmosphere are wrong and their abundances below the clouds are within a factor of two of each other, or that some alternative chemistry explains the \ce{SO_2} depletion. We are not the first to notice the implications of this puzzle: a missing sulfur reservoir \citep[see, e.g.][]{Yung1982,Parkinson2015,Marcq2018}.

One alternative mechanism to explain the \ce{SO_2} depletion is the formation of condensible sulfur allotropes out of \ce{SO_2} photodissociation or thermal dissociation products. The remaining sulfur cannot be in the form of \ce{SO}, because the resulting concentrations of \ce{SO} would be at least two orders of magnitude greater than indicated by above-cloud observations. This explanation requires photodissocation of both \ce{SO} and \ce{SO_2} near the cloud top that is many orders of magnitude more efficient than predicted by any model, or similarly more efficient thermal dissociation of \ce{SO_2}, and must explain 80\% of the \ce{SO_2} depletion. 20\% of the \ce{SO_2} will be converted into \ce{SO_3} and will react with \ce{H_2O} to form \ce{H_2SO_4}, condensing out. Another 20\% will balance the reducing power of the \ce{SO_3} removal. The remaining 60\% would have to go through the either or both of the total reactions:
\begin{align}
2\ce{SO_2} &\rightarrow \ce{S_2} + 2 \ce{O_2}, & {\rm Thermochemistry};\\
2\ce{SO_2} + 4h\nu &\rightarrow \ce{S_2} + 2 \ce{O_2} &  {\rm Photochemistry}.
\end{align}
This would either predict 150 ppm additional \ce{O_2} in the upper atmosphere of Venus, exacerbating the \ce{O_2} overabundance problem discussed in the Introduction, or would lead to oxidation of \ce{CO}. However, oxidation of \ce{CO} would cause it to become depleted, whereas we see that \ce{CO} increases above the clouds. It is not possible that the excess oxygen would remain in the form of atomic oxygen, because atomic oxygen is not chemically stable at above-cloud altitudes. It is possible that the oxygen is stored in some other chemical species that has not yet been identified, but thus far there is no known candidate species at several ppm concentrations needed to contain the excess oxygen. For these reasons we do not consider this explanation further here.

\section{The Model} 
\label{sec:model}

For this work we use a photochemistry-diffusion model based on the model of \citet{Rimmer2016}. The model is composed of a solver and a network. The solver, \textsc{Argo}, solves the time-dependent set of coupled non-linear differential equations:
\begin{equation}
\dfrac{dn_{\ce{X}}}{dt} = P_X - L_Xn_{\ce{X}} - \dfrac{\partial \Phi_X}{\partial z},
\end{equation}
where $n_{\ce{X}}$ (cm$^{-3}$) is the number density of species $\ce{X}$ at height $z$ (cm) and time $t$ (s), $P_X$ (cm$^{-3}$ s$^{-1}$) is the rate of production of \ce{X} at height $z$ and time $t$, $L_X$ (s$^{-1}$) is the rate constant for destruction of species \ce{X} at height $z$ and time $t$. The term $\partial \Phi_X/\partial z$ (cm$^{-3}$ s$^{-1}$) describes the divergence of the vertical diffusion flux.

As described by \citet{Rimmer2016}, the chemistry is solved by following the motion of a parcel up and down through a one-dimensional atmosphere described by a grid of set temperature, $T$ (K), pressure, $p$ (bar) and other properties. A parcel starts at the surface with a particular set of initial chemical conditions, and then moves through the atmosphere at a velocity determined by the Eddy diffusion coefficient:
\begin{equation}
t_{\rm chem} = \dfrac{\big(\Delta z\big)^2}{2 K_{zz}},
\end{equation}
where $\Delta z$ (km) is the change in height from one part of the grid to the next, and $K_{zz}$ (cm$^2$ s$^{-1}$) is the Eddy diffusion coefficient. The volume mixing ratios are recorded for each species at each grid height, constructing chemical profiles for the atmosphere. There is a method to account for molecular diffusion, described by \citet{Rimmer2016} and corrected by \citet{Rimmer2019Err}. We do not solve above Venus's homopause, so molecular diffusion, though included, is not significant and we do not describe it in detail here. After the parcel makes a single complete trip, a UV radiative transfer model is run on the recorded profiles, completing a single global iteration and recording the actinic flux\footnote{The actinic flux is the total (direct and diffusive) spectral irradiance integrated over a unit sphere.} $F_{\lambda}$ (photons cm$^{-2}$ s$^{-1}$; photons will be excluded from the units hereafter).

This method reproduces results for modern Earth and Jupiter \citep{Rimmer2016} and agrees with Eulerian solvers for chemical quenching heights in Hot Jupiter atmospheres \citep{Tsai2017,Hobbs2019,Hobbs2020}. 

Some of the values of $L_X$ (and corresponding production coefficients) are the result of direct photodissociation and photoionization. These are all set to zero for the first global iteration, and then are calculated for global iteration $I$ using the chemical profiles of global iteration $I-1$. The photodissociation and photoionization rate constant for species \ce{X} are:
\begin{equation}
k_{\lambda}(\ce{X}) = \dfrac{1}{2}\int F_{\lambda} \, \sigma_{\lambda}(\ce{X}) \; d\lambda
\end{equation}
where $\lambda$ (\AA) is the wavelength, $z$ (km) is the atmospheric height, $\sigma_{\lambda}(\ce{X})$ (cm$^2$) are the photochemical cross-sections (see Appendix \ref{app:stand}). The factor of 1/2 is typically included to account for rotation of the planet. Venus rotates too slowly to include this factor for the same reason, and would typically be treated as having a dayside and nightside chemistry. However, the zonal winds of Venus are very strong, horizontal mixing is fast, and for simplicity we consider the atmosphere to be a well-mixed average of day-side and night-side chemistries, which will be true for the long-lifetime and medium-lifetime species. The flux at height $z$ is given by:
\begin{equation}
F_{\lambda}(z) = F_{\lambda}(z_0) e^{-(\tau + \tau_a)/\mu_0} + F_{\rm diff}.
\label{eqn:actinic-flux}
\end{equation}
Here $F_{\lambda}(z_0)$ (cm$^{-2}$ s$^{-1}$ \AA$^{-1}$) is the top-of-atmosphere (TOA) flux as described in Section \ref{sec:model-initial}, $F_{\rm diff}$ (cm$^{-2}$ s$^{-1}$ \AA$^{-1}$) is the diffuse flux from scattering \citep[see][]{Rimmer2016}, the cosine of the average zenith angle $\mu_0 = 0.54$ \citep[see][]{Hu2012}, $\tau$ is the optical depth from molecular absorption calculated from the chemical profiles using the prior global solution as well as the photochemical cross-sections (see Appendix \ref{app:stand} for details). In addition, $\tau_a$ is the additional optical depth due to Venus's mysterious UV absorber, described in Section \ref{sec:model-initial}.

Beyond these photodissociation and photoionization rate constants, the coefficients used to construct $P_X$ and $L_X$ are provided by the chemical network, \textsc{Stand2020}, which we introduce here. We start with the sulfur network of \citet{Hobbs2020} and add all relevant reactions involving species without thermochemical data from \citet{Greaves2020}. The network includes 2901 reversible reactions and 537 irreversible reactions involving 480 species comprised of H/C/N/O/S/Cl and a handful of other elements, including condensation of \ce{H_2O}, and a host of other species, most of which condense at temperatures far lower than achieved in Venus's atmosphere, see Appendix \ref{app:stand}. The network, including added condensation chemistry for sulfuric acid and sulfur allotropes, is described in detail in Appendix \ref{app:stand}.

In addition, we track the dissolution of \ce{SO_2} into the cloud droplets and subsequent liquid-phase chemistry for the cloud chemistry model described in Section \ref{sec:cloud}. 

The chemical profiles from the most recent global iteration, $I$ are compared to the profiles from the next most recent global iteration, $I-1$ in order to determine convergence. Convergence criteria are the same as for \citet{Greaves2020}.

We give the parameters and initial conditions for our model in Section \ref{sec:model-initial}, including the temperature profile, Eddy diffusion, stellar irradiation, and surface boundary conditions for the chemistry. In Section \ref{sec:equilib}, we discuss the consistency of these initial conditions compared to chemical equilibrium at the surface.

\subsection{Parameters and Initial Conditions}
\label{sec:model-initial}

\begin{figure}[ht!]
\plottwo{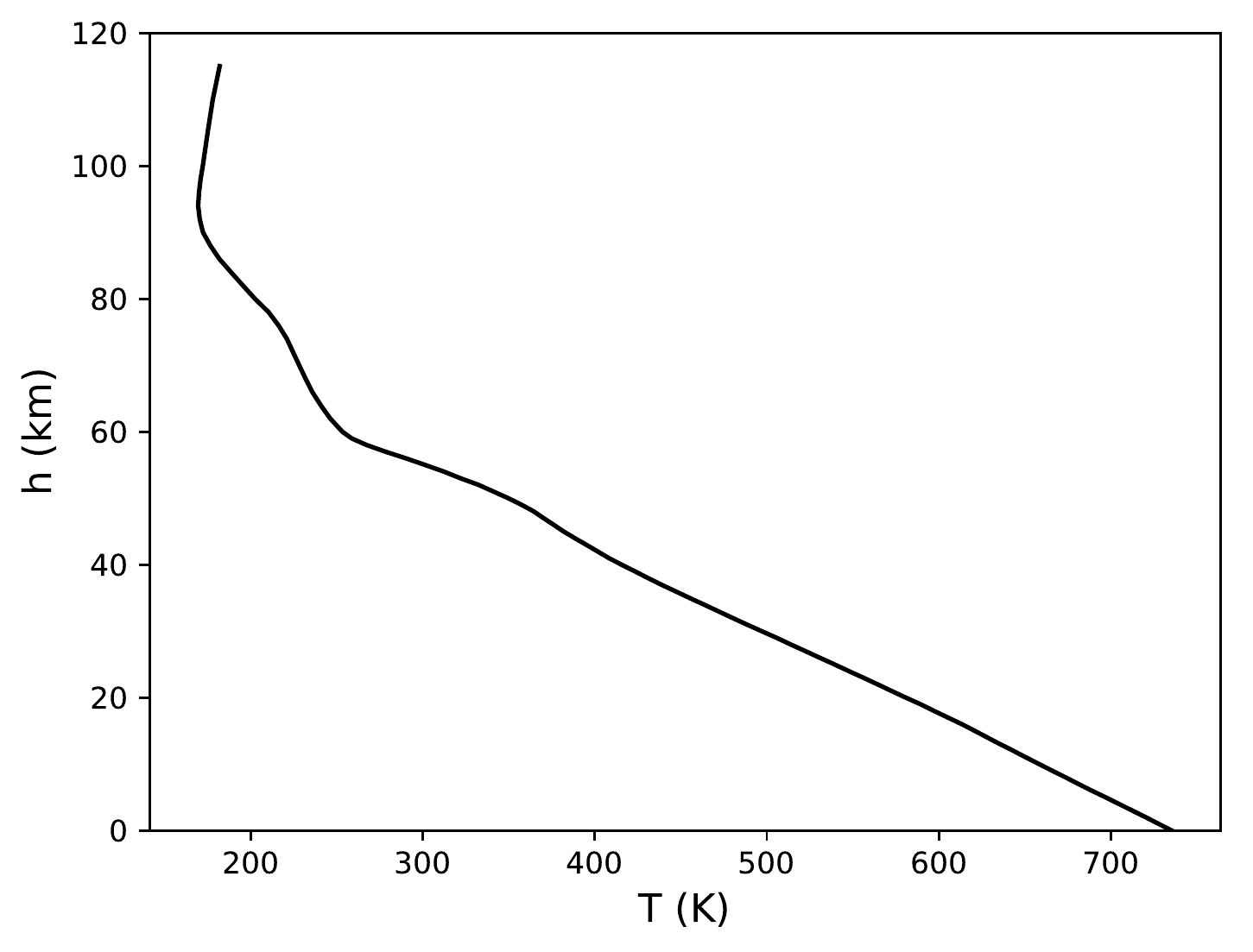}{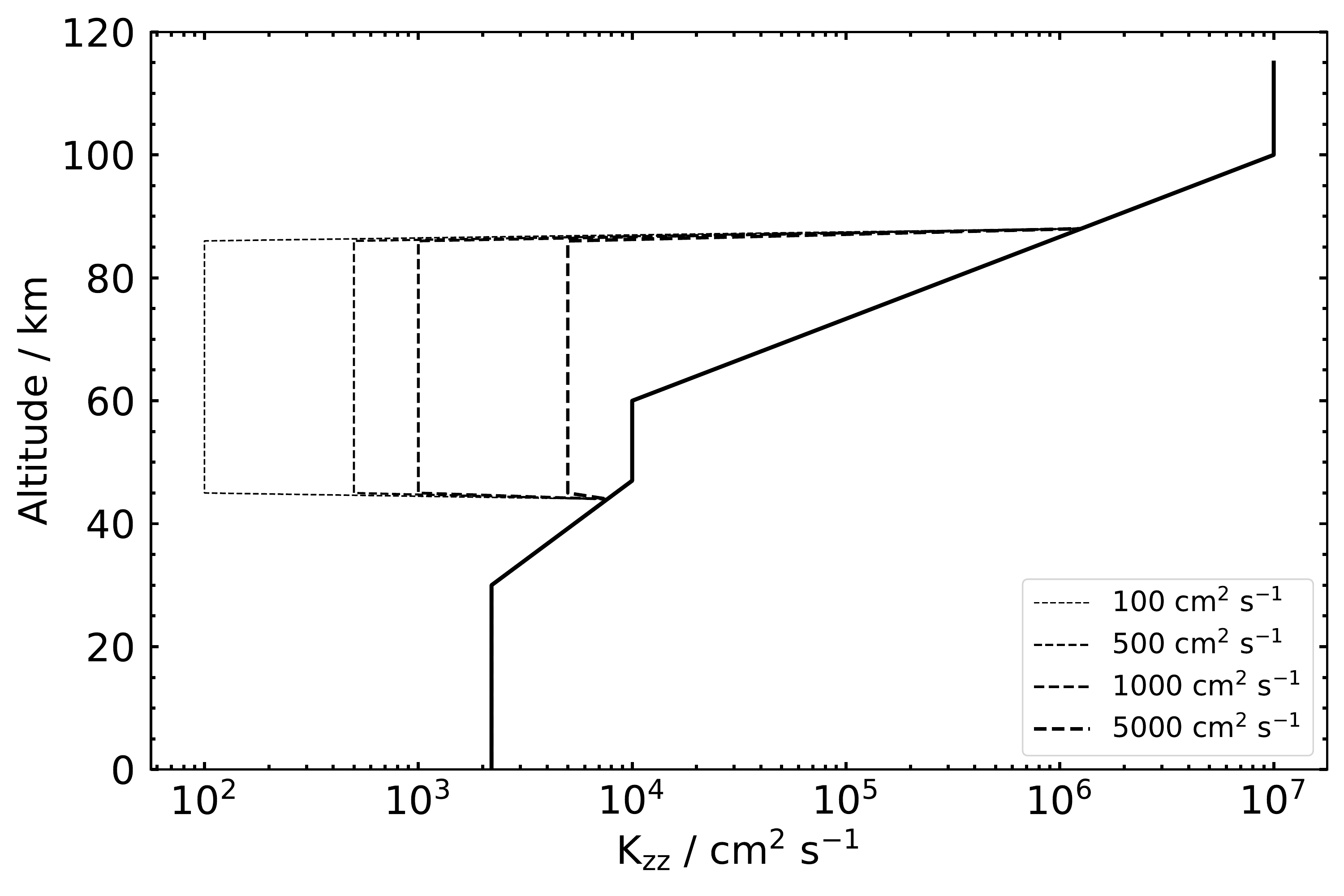}
\caption{Atmospheric temperature, $T$ (K, left), and $K_{zz}$ (cm$^{2}$ s$^{-1}$, right), as a function of height $h$ (km) \citep[from][]{Kras2007}. The two values of $K_{zz}$ are for the cloud chemistry and the low-sulfur models (Sections \ref{sec:abundance}, \ref{sec:cloud}, \ref{sec:results-so2} and \ref{sec:results-clouds}, solid)  and high-water models (Sections \ref{sec:abundance} and \ref{sec:results-h2o}, dashed and dotted), from \citet{Kras2007,Kras2012} and \citet{Bierson2020} respectively. \label{fig:profiles}}
\end{figure}

The initial conditions and parameters we use are are similar to those used for \citet{Greaves2020}. We use the same fixed temperature profile as \citet{Greaves2020}, which was initially taken from \citet{Kras2007} and \citet{Kras2012}. We use Eddy diffusion profiles from the same sources, though we also explore the effect of using the in-cloud Eddy diffusion coefficients of \citet{Bierson2020} in Section \ref{sec:abundance}. The profiles we use are shown in Figure \ref{fig:profiles}.

We use a scaled top-of-atmosphere (TOA) solar spectrum from 1 -- 10000 \AA, compiled by \citet{Granville2017}, for our TOA boundary condition. This data was compiled using \citet{Matthes2017} for the 401 -- 1149 \AA\ spectral region and \citet{Coddington2015} for the other wavelengths. The actinic flux is then multiplied by 1.913 to account for the difference in average distances of Earth and Venus from the Sun. The TOA spectrum is included in the Supplementary Materials.

We also include, in addition to molecular absorption and scattering as described above in Section \ref{sec:model}, a parameterization of the mysterious UV absorber present within Venus's cloud layer. The parameterization originates with \citet{Kras2012} and is the same as that used by \citet{Greaves2020}, it takes the form:
\begin{align}
\dfrac{d\tau_a}{dz} &= 0.056 \; \mathrm{km}^{-1} \exp\Big\{-\dfrac{z-67 \, \mathrm{km}}{3 \, \mathrm{km}} - \dfrac{\lambda - 3600 \mathrm{\AA}}{1000 \mathrm{\AA}}\Big\}, & z > 67 \; \mathrm{km}; \notag\\
\dfrac{d\tau_a}{dz} &= 0.056 \; \mathrm{km}^{-1} \exp\Big\{- \dfrac{\lambda - 3600 \mathrm{\AA}}{1000 \mathrm{\AA}}\Big\}, & 58 \; \mathrm{km} < z \leq 67 \; \mathrm{km}; \label{eqn:mysterious-flux}\\
\dfrac{d\tau_a}{dz} &= 0.0 \; \mathrm{km}^{-1}, & z \leq 58 \; \mathrm{km}. \notag
\end{align}
Where $\tau_a$ is added to the optical depth in Eq. (\ref{eqn:actinic-flux}).

The initial surface mixing ratios we set for Venus's atmosphere are given in Table \ref{tab:init}, defining the bulk atmosphere. For all of the models presented outside of Section \ref{sec:ph3}, we do not include \ce{PH_3}. Now that we've defined our model and parameters, we will lay out the different scenarios that we consider for explaining the observed \ce{SO_2} depletion.

\begin{deluxetable*}{cccccccccc}
\tablecaption{Initial Surface Abundances used for Model Atmospheres of Venus\label{tab:init}}
\tablehead{
\colhead{\ce{CO_2}} & \colhead{\ce{N_2}} & 
\colhead{\ce{SO_2}} & \colhead{\ce{H_2O}} & 
\colhead{\ce{OCS}} & \colhead{\ce{CO}} & \colhead{\ce{HCl}} & \colhead{\ce{H_2}} & \colhead{\ce{H_2S}} & \colhead{\ce{NO}}
} 
\startdata
0.96 & 0.03 & 150 ppm$^*$ & 30 ppm$^*$ & 5 ppm & 20 ppm & 500 ppb & 3 ppb & 10 ppb & 5.5 ppb\\
\enddata
\vspace{1mm}
$^*$The mixing ratios of \ce{SO_2} and \ce{H_2O} are varied for some of the models.
\end{deluxetable*}

\subsection{Equilibrium Surface Composition}
\label{sec:equilib}

Here we discuss the consistency of the chemical boundary conditions (Table \ref{tab:init}) compared to chemical equilibrium, and the implications these conditions may have on surface mineralogy. We find there are solutions where the gas-phase chemistry is consistent with our chosen boundary conditions and the condensed-phase chemistry is broadly consistent with observed surface mineral compositions. We will explore these conditions and their implications for hypothetical cloud chemistry in Section \ref{sec:discussion}.

In June 1985, the Vega 2 lander determined the composition of the Venusian surface rock in the northern region of Aphrodite Terra \citep[][see also \citealt{Fegley2014}]{Surkov1986}.  The rock was analysed by X-ray fluorescence employing instrumentation that had been improved based on the experience with the previous Venera 13 and 14 missions \citep{Surkov1982,Surkov1984}. The measured oxide ratios are listed in Table~\ref{tab:Vega2}. We have used these data, in combination with the observed gas composition at the surface in Table~\ref{tab:init}, to investigate the question in how far the gas at the bottom of the Venusian atmosphere is in chemical equilibrium and in phase equilibrium with the surface rock.

\begin{table}[!b]
\vspace*{4mm}
\caption{Oxide mass fractions [\%] of the surface rock measured by the
  Vega 2 lander \citep{Surkov1986}.}
\label{tab:Vega2}
  \hspace*{-30mm}\resizebox{21.2cm}{!}{
  \begin{tabular}{cccccccccc}
  \hline
  \hline
  &&&&&&&&&\\*[-3.4ex]  
  \ce{SiO2} & \ce{TiO2} & \ce{Al2O3} & \ce{FeO} & \ce{MnO} & \ce{MgO} &
  \ce{CaO} & \ce{Na2O} & \ce{K2O} & \ce{SO3}\\
  $45.6 \pm 3.2$ & $0.2 \pm 0.1$ & $16 \pm 1.8$ & $7.7 \pm 1.1$ & 
  $0.14 \pm 0.12$ & $11.5 \pm 3.7$ & $7.5 \pm 0.7$ & $2.0$ & 
  $0.1\pm 0.08$ & $4.7 \pm 1.5$\\
  \hline
  \end{tabular}}\\[1mm]
  \hspace*{-2mm}\resizebox{18.3cm}{!}{
  \footnotesize{The following elements were not detected: 
  Cl, Cu and Pb $<$0.3\% , Zn$<$0.2\%, Pb, As, Se, Br, Sr, Y, Zr, Nb, Mo $<$0.1\%}.}
\end{table}

These investigations were carried out by means of the chemical and phase equilibrium code {\sc GGchem} \citep{Woitke2018} taking into account the following 16 elements: H, C, N, O, F, S, Cl, Fe, Mn, Si, Mg, Ca, Al, Na, K, and Ti.  No information is available about phosphorous at the Venusian surface, so we have excluded that element from this investigation.  {\sc GGchem} finds 442 gas phase species (atoms, ions, molecules and molecular ions) and 190 condensed species in its databases for this selection of elements. The thermochemical data for the molecules are based on the NIST-Janaf tables \citep{Chase1982, Chase1986, Chase1998}, as fitted by \citet{Stock2008}, with some additions for diatomic molecules from \citet{Barklem2016}.  Condensed phase data are taken from the SUPCRTBL database \citep{Zimmer2016} and from the NIST-Janaf database. Some additional vapour pressure data are taken from \citet{Yaws1999}, \citet{Weast1971}, \citet{Ackerman2001}, and \citet{Zahnle2016}.

We consider a mixture of gas and condensed species at $T\!=\!735\,$K and $p\!=\!90\,$bars with total (gas\,$+$\,condensed) element abundances $\epsilon^0$, see \citet[][their figure~1]{Herbort2020}. In order to find these element abundances we first convert the solid oxide mass ratios given in Table~\ref{tab:Vega2} into element particle ratios. Second, we multiply by an arbitrary factor of 1000 and then add the observed gas phase element abundances from Table~\ref{tab:init}.  Third, we carefully adjust the total oxygen abundance $\epsilon^0_{\rm O}$ until the gas over the condensates has a \ce{SO2} concentration of 150\,ppm in the model. The arbitrary factor in preparation step two causes the model to produce mostly condensed phases with only small amounts of gas above it. That factor has little influence on the results as long as it is large. The reason for this behaviour is that once {\sc GGchem} has determined the solid composition in form of active condensates, which have supersaturation ratio $S\!=\!1$ (all other condensates have $S\!<\!1$), one can add arbitrary amounts of those condensates to $\epsilon^0_{\rm O}$, they will just fall out again without changing the resulting gas composition.

The results of this model are shown in Table~\ref{tab:GGchem}. The resultant solid composition of the Venusian surface rock is a felsic mixture of enstatite (\ce{MgSiO3[s]}) and quartz (\ce{SiO2[s]}).  The condensates anorthite (\ce{CaAl2Si2O8[s]}), albite (\ce{NaAlSi3O3[s]}), and microcline (\ce{KAlSi3O8[s]}) are the three main minerals forming feldspar which is found e.g.\ in basaltic rock on Earth. Iron is found to be entirely bound in magnetite (\ce{Fe2O3[s]}). No carbonates and no phyllosilicates are found to be stable under the assumed conditions, nor any minerals containing chlorine. The only halide found to be stable is magnesium fluoride (\ce{MgF2[s]}). Therefore, all carbon, nitrogen, hydrogen and chlorine assumed in the model is present in the gas, which allows us to directly fit the observed gas phase concentrations of \ce{CO2}, \ce{N2}, \ce{H2O} and \ce{HCl}.

Fitting the \ce{SO2} concentration is more difficult, because sulphur is mostly contained in anhydrite (\ce{CaSO4}).  In the close vicinity of the equilibrium solution outlined in Table~\ref{tab:GGchem}, we see that additional oxygen is used to form more anhydrite in the model on the expense of gaseous \ce{SO2} and anorthite (\ce{CaAl2Si2O8[s]}) via
\begin{equation}
  \ce{O} \,+\, \ce{SO2} \,+\, \ce{CaAl2Si2O8[s]} ~\longleftrightarrow~
  \ce{CaSO4[s]} \,+\, \ce{Al2O3[s]} \,+\, 2 \ce{SiO2[s]} \ ,
  \label{eq:regulation}
\end{equation}
which is a potentially important buffer mechanism to understand the \ce{SO2} concentration in the lower Venus atmosphere.  It allows us in the model to adjust the total oxygen abundance $\epsilon^0_{\rm O}$ to find the desired \ce{SO2} concentration (more oxygen means less \ce{SO2}).

Table~\ref{tab:GGchem} shows that it is possible to explain both the observed composition of the near-crust Venusian atmosphere and the solid composition of the surface rock by a simple consistent phase equilibrium model. All molecules that are predicted to be abundant in our model (those with percent or ppm concentrations) have observed counterparts.  The abundance hierarchy matches between model and observations.  Other common molecules like \ce{CH4} and \ce{NH3}, which have extremely small abundances ($<\!10^{-15}$) in our model, are not observed. The Venus atmosphere can hence be classified as type~B atmosphere according to \citet{Woitke2020}, along with the atmospheres of Earth and Mars.  All predicted molecular concentrations are in reasonable agreement with the observed values, in particular when taking into account the large measurement uncertainties (see Table~\ref{tab:obs}).  We note, however that this does not prove that the Venusian atmosphere is in chemical and phase equilibrium with the surface rock, it only shows that the data can be interpreted that way. 

For the water-rich scenario, see Sect.~\ref{sec:abundance}, we can increase the hydrogen abundance to find a model with 200\,ppm \ce{H2O}, which has little effect on CO and OCS, but results in  slightly increased abundances of the ppb molecules,
149\,ppm \ce{SO2},
12\,ppm \ce{CO},
9\,ppm \ce{OCS},
505\,ppb \ce{HCl},
505\,ppb \ce{HF},
110\,ppb \ce{S2},
370\,ppb \ce{H2S},
110\,ppb \ce{S2O},
and 18\,ppb \ce{H2}, which is arguably still consistent with the observations. 
For the sulphur-poor scenario, we can increase the oxygen abundance to find a model with 20\,ppm \ce{SO2}. In that case, the atmosphere is found to have a purer, more oxidising character with
30\,ppm \ce{H2O},
2\,ppm \ce{CO},
4\,ppb \ce{OCS},
505\,ppb \ce{HCl},
355\,ppb \ce{HF},
and \ce{S2}, \ce{H2S}, \ce{S2O}, \ce{H2} all $<$\,1ppb,
which seems inconsistent with the observations.

\begin{table}
\caption{Results of the {\sc GGchem} model for the bottom of the
  Venusian atmosphere assuming the gas to be in chemical equilibrium 
  and in phase equilibrium with the surface rock at $T\!=\!735\,$K and
  $p\!=\!90\,$bars.}
\label{tab:GGchem}
{\bf Gas phase composition}\\[1mm]
\hspace*{-38.5mm}\resizebox{22cm}{!}{\begin{tabular}{c|cccccccccccc}
  \hline
  \hline
  &&&&&&&&&\\*[-3.4ex]
  &
  \ce{CO2} & \ce{N2 } & \ce{SO2} & \ce{H2O} & \ce{CO } & \ce{OCS} &
  \ce{HCL} & \ce{HF } & \ce{S2 } & \ce{H2S} & \ce{S2O} & \ce{H2
  }\\
  \hline
  &&&&&&&&&\\*[-3.3ex]
  input &
  96\,\% & 3\,\% & 150\,ppm & 30\,ppm & 20\,ppm & 5\,ppm &
  500\,ppb & 500\,ppb & - & 10\,ppb & - & 3\,ppb\\
  result &
  97\,\% & 3\,\% & 150\,ppm & 30\,ppm & 12\,ppm & 9\,ppm &
  505\,ppb & 355\,ppb & 114\,ppb & 57\,ppb & 17\,ppb & 3\,ppb\\
  \hline
\end{tabular}}\\[1mm]
{\footnotesize All other molecules have concentrations $<$\,1\,ppm in
  the model.}\\[2mm]
{\bf Solid composition} (results in mass fractions)\\[1mm]
\hspace*{-32mm}\resizebox{21.4cm}{!}{\begin{tabular}{ccccccccccc}
  \hline
  \hline
  &&&&&&&&&\\*[-3.4ex]
  \ce{MgSiO3} & \ce{SiO2} & \ce{CaAl2Si2O8} & \ce{NaAlSi3O8} &
  \ce{CaSO4} & \ce{Fe2O3} &
  \ce{Al2O3} & \ce{TiO2} & \ce{KAlSi3O8} & \ce{Mn3Al2Si3O12} & \ce{MgF2}\\
  &&&&&&&&&\\*[-3.3ex]
  29.7\,\% & 7.6\,\%  & 21.7\,\%  & 17.6\,\% & 8.3\,\% & 8.9\,\% &
  5.1\,\% & 0.2\,\% & 0.6\,\% & 0.3\,\% & (trace) \\
  \hline
\end{tabular}}\\[1mm]
\footnotesize{\ce{MgSiO3} $=$ enstatite,
  \ce{SiO2}  $=$ quartz,
  \ce{CaAl2Si2O8}  $=$ anorthite,
  \ce{NaAlSi3O8}   $=$ albite,
  \ce{CaSO4}  $=$ anhydrite,\\
  \ce{Fe2O3}   $=$ magnetite,
  \ce{Al2O3}  $=$ corundum,
  \ce{TiO2}  $=$ rutile,
  \ce{KAlSi3O8}  $=$ microcline,
  \ce{Mn3Al2Si3O12}  $=$ spessartine,\\
  \ce{MgF2}  $=$ magnesium fluoride.
  All other condensates are under-saturated in the model and have zero mass fractions.}
\end{table}

\section{Hypothesis: The Observational Constraints are Wrong} 
\label{sec:abundance}

One possibility to consider is that the below-cloud observational constraints are incorrect. Possibly the below-cloud water vapor is much higher than observations suggest, a possibility considered by \citet{Yung1982}. Or the below-cloud sulfur dioxide is much lower than most observations suggest.

In order to explore the sulfur-poor hypothesis we vary the below-cloud \ce{SO_2} abundance from 80 ppm down to 6 ppm with below-cloud \ce{H_2O} kept at the nominal value (30 ppm). To explore the water-rich hypothesis we vary the below-cloud \ce{H_2O} abundance from 30 ppm up to 200 ppm with \ce{SO2} kept at the nominal value (150 ppm). In this case we find that the observed above-cloud \ce{SO2} depletion is not achieved for any value of below-cloud \ce{H_2O} due to \ce{H_2O} self-shielding effects. Some support for the water-rich hypothesis is the $>100 \, {\rm ppm}$ abundances of water vapor observed in and directly below the clouds of Venus \citep{Mukhin1982,Surkov1982,Bell1991}, though these are inconsistent with $<100 \, {\rm ppm}$ abundances close to the surface \citep{Bertaux1996}.

To explore the possibility of the water-rich hypothesis further we test the effect of introducing a trap in the eddy diffusion profile within the cloud layer alongside varying the \ce{H_2O} abundance below the clouds. Observational constraints on the eddy diffusivity as a function of altitude in the atmosphere are sparse. \citet{Marcq2018} have suggested that the existence of statically stable layers in the cloud region may inhibit dynamical exchange between the upper and lower regions of the atmosphere, a possibility explored by \citet{Bierson2020}. In the present work we test this possibility by modifying the nominal eddy diffusion profile taken from \citet{Kras2007} and \citet{Kras2012} to include a trap of constant lower $K_{zz}$ across the extent of the cloud layer. The range of values that we explore for $K_{zz}$ in the trap are 5000, 1000, 500 and $100 {\rm \, cm^2 \, s^{-1}}$, shown in figure \ref{fig:profiles}. Such a $K_{zz}$ trap, if it exists in the Venus cloud layer, is particularly relevant to the water-rich hypothesis as enhanced water abundance below the clouds would result in lesser thermal heating flux at the cloud base due to \ce{H_2O} IR absorption, which in turn increases the convective stability and decreases the eddy diffusivity, shown by \citet{Yamamoto2014}. We investigate the results from combining a $K_{zz}$ trap and enhanced below-cloud water abundance.

\section{Hypothesis: Another Source of Hydrogen in the Clouds} 
\label{sec:cloud}

The \ce{SO_2} depletion can be explained if there is another source of hydrogen in the clouds. Here we will use \ce{NaOH} as that source of hydrogen for the sake of convenience when calculating the cloud droplet chemistry. We are not claiming that this is the source of hydrogen. We will discuss possible sources of this excess hydrogen in Section \ref{sec:discussion}.

Some \ce{SO_2} will dissolve into the cloud droplets directly, with a concentration in the droplet, $c {\rm \,(mol/L)}$, linearly proportional to the partial pressure of \ce{SO_2}, with the Henry's Law constant, $H_{\ce{SO_2}} {\rm \big(mol/(L bar)\big)}$, as the constant of proportionality: 
\begin{equation}
c(\ce{SO_2}) = H_{\ce{SO_2}} \; p f_{\ce{SO_2}}.
\label{eqn:henrys-law}
\end{equation}
Here $p f_{\ce{SO_2}}$ is the partial pressure of \ce{SO_2}, with the total gas pressure $p$ (bar) and $f_{\ce{SO_2}}$ as the volume mixing ratio of \ce{SO_2}, equal to $n_{\ce{SO_2}}/\big(\Sigma_{\ce{X}} n_{\ce{X}}\big)$.

For \ce{SO_2} dissolved in water, the Henry's Law constant is $\approx 10^{-2}$ mol/(L bar) \citep{Burkholder2020}. For \ce{SO_2} dissolved in sulfuric acid, the constant increases with the sulfuric acid concentration between 0.1 and 1 mol/(L bar) \citep{Zhang1998}. The \ce{SO_2} then participates in the following reactions (here g is gas-phase, $\ell$ is in the droplet). We first consider the dissolution of \ce{SO_2} and \ce{H_2O} into the droplets by Henry's law.
\begin{align}
\ce{SO_2(g)} &\rightleftharpoons \ce{SO_2($\ell$)}, \label{eqn:liquid-first}\\
\ce{H_2O(g)} &\rightleftharpoons \ce{H_2O($\ell$)}. \label{eqn:liquid-second}
\end{align}
The rate constants for this reaction are balanced such that the concentration of \ce{SO_2} at any point agrees with Eq. (\ref{eqn:henrys-law}) when accounting for the droplet volume (see below). Also, in reality, the Henry's law constant for both water vapor and sulfur dioxide will vary with the composition of the droplet, and the composition of the droplet will vary as more water vapor and sulfur dioxide dissolves in the droplet. Our model does not account for these variations, and thus Henry's law as treated here is only an approximation. The rest of the reactions are dissociation reactions for which bimolecular rate constants are set to $k_f = 5 \times 10^{10}\,{\rm mol^{-1} \, L \, s^{-1}}$ in order to rapidly achieve equilibrium and avoid any dynamic effects, and the unimolecular rate constant is assigned a value that preserves equilibrium set by the $pK_a$ or $pK_b$. The rate constant for the reverse reaction is (with units s$^{-1}$):
\begin{align}
k_r &= \dfrac{\rho k_f}{\mu} \, 10^{-pK_a}, \\
k_r &= \dfrac{\rho k_f}{\mu} \, 10^{-pK_b},
\end{align}
where $\rho\,{\rm (g\,cm^{-3})}$ is the density of the liquid and $\mu\,({\rm g\, mol^{-1}})$ is the molar weight of the species. The reactions for the sulfates are:
\begin{align}
\ce{H_2SO_4($\ell$)} &\rightleftharpoons \ce{HSO_4^-($\ell$)} + \ce{H^+($\ell$)}, & pK_{a,1} = -2.8;\\
\ce{HSO_4^-($\ell$)} &\rightleftharpoons \ce{SO_4^{2-}($\ell$)} + \ce{H^+($\ell$)}, & pK_{a,2} = 1.99;
\end{align}
and the reactions involving sulfurous acid and sulfites:
\begin{align}
\ce{SO_2($\ell$)} + \ce{H_2O($\ell$)} &\rightleftharpoons \ce{HSO_3^-($\ell$)} + \ce{H^+($\ell$)}, & {\rm see \; below};\\
\ce{H_2SO_3($\ell$)} &\rightleftharpoons \ce{SO_2($\ell$)} + \ce{H_2O($\ell$)}, & {\rm see \; below};\\
\ce{H_2SO_3($\ell$)} &\rightleftharpoons \ce{HSO_3^-($\ell$)} + \ce{H^+($\ell$)}, & pK_{a,1} = 1.857;\\
\ce{HSO_3^-($\ell$)} &\rightleftharpoons \ce{SO_3^{2-}($\ell$)} + \ce{H^+($\ell$)}  & pK_{a,2} = 7.172.
\end{align}
The rate constant for \ce{SO_2} to react with \ce{H_2O} is $2 \times 10^8 \, {\rm mol^{-1} \, L \, s^{-1}}$ \citep{Eigen1961,Brandt1995} and the rate constant for \ce{H_2SO_3} dissociation is $10^8$ s$^{-1}$ \citep{Eigen1961,Brandt1995}. These reactions are sufficiently fast to draw the gas-droplet chemistry into equilibrium at all atmospheric heights. It was for this reason that we chose \ce{NaOH} as our candidate salt. Finally, we consider the self-dissociation of water and the dissociation of sodium hydroxide:
\begin{align}
\ce{H^+($\ell$)} + \ce{OH^-($\ell$)} &\rightleftharpoons \ce{H_2O($\ell$)}, & pK_a = 14\\
\ce{NaOH($\ell$)} &\rightleftharpoons \ce{Na^+($\ell$)} + \ce{OH^-($\ell$)}, & pK_b = 0.2 \label{eqn:liquid-last}
\end{align}
Sodium hydroxide is the example species we will use to buffer the clouds of Venus, freeing up more water in the droplet to react with \ce{SO_3}, forming sulfuric acid, or with \ce{SO_2} to form sulfurous acid.

If the excess source of hydrogen is a salt, as is the case with \ce{NaOH}, \ce{H^+} is replaced by \ce{Na^+}. We calculate the pH where $\ce{pH} = - \log_{10}\big(a_{\mathrm{H^+}}\big)$, where $a_\mathrm{H^+}$ is the \ce{H^+} activity. The concentration of \ce{NaOH} needed in the droplets to sequester \ce{SO_2} is determined by considering the cloud droplet volume as a function of height:
\begin{equation}
V_d(h) = \int \dfrac{4}{3}\pi r_d^3 \dfrac{\partial N_d}{\partial r_d} \, dr_d,
\label{eqn:droplet-volume}
\end{equation}
where $r_d$ ($\mu$m) is the droplet radius, and the function $r_d \big(\partial N_d/\partial r_d\big)$ is the droplet size distribution, which we take from \citet{Gao2014}:
\begin{equation}
\dfrac{\partial N_d}{\partial r_d} = \dfrac{\partial n_d}{\partial r_d} V_{\rm atm}
\end{equation}
where $n_d$ (cm$^{-3}$) is the droplet number density (across all sizes) and $V_{\rm atm} = 4\pi R_p^2 \Delta z$, with $R_p = 6052$ km as the radius of Venus and $\Delta z$ (km) is the model height step.  We can calculate the amount of \ce{NaOH} needed to deplete the \ce{SO2} to the observed levels. We do this by dividing the total number of \ce{SO_2} molecules by the total droplet volume at height $z$:
\begin{equation}
c(z) = \dfrac{p f_{\ce{SO_2}}(z)}{N_AkT(z)} \, \Bigg[\int \dfrac{4}{3}\pi r_d^2 \Bigg(r_d \dfrac{\partial n_d(z)}{\partial r_d}\Bigg)\, dr_d\Bigg]^{-1},
\end{equation}
where $N_A = 6.022 \times 10^{23}$ is Avogadro's Number and $k = 1.38065 \times 10^{-16}$ erg/K is Boltzmann's constant. The above equation is used to prescribe the initial concentration of \ce{NaOH} (in units of mol/cm$^3$) for our solver. The \ce{NaOH} will rapidly dissociate and the \ce{OH^-} will then recombine with \ce{H^+} to form \ce{H_2O}, and this, along with dissolved \ce{H_2O} from the gas phase will react with \ce{SO_2} to form \ce{H_2SO_3}. This will rapidly dissociate to form \ce{HSO_3^-} and \ce{H^+}, and will buffer the solution, meaning that if the \ce{H^+} activity is perturbed, the reactions between the anions and \ce{H^+} will bring the activity of \ce{H^+} back to a given value determined by the $pK_a$. It should be emphasized that Eq's (\ref{eqn:liquid-first}) and (\ref{eqn:liquid-second}) do not describe condensation of either of these species, who generally have partial pressures throughout Venus's atmosphere that are well below the vapor pressure. Rather, these are the equilibrium concentrations in the droplets due to dissolution, balanced with the partial pressures in the gas.

Prescribing the \ce{NaOH} this way introduces extra hydrogen into the model in a way not accounted by mass balance or atmospheric redox balance. A full solution of the atmospheric chemistry coupled with surface chemistry would preserve this balance, but that would require us to identify the source of hydrogen. Though we speculate on possible sources of hydrogen, from the surface (delivered via volcanism or winds) or exogenous, in Section \ref{sec:discussion}, we do not know enough about Venus's atmosphere, surface or clouds to confidently identify a source, and therefore do not self-consistently include this source in our model.

\section{Results} 
\label{sec:results}

{\bf The depletion of \ce{SO_2} can be explained by the \ce{SO_2} dissolving into the clouds, if the cloud pH is higher than previously believed.} It can also be explained by varying the below-cloud \ce{SO_2} and \ce{H_2O}. We discuss the consequences of varying the \ce{SO_2} in Section \ref{sec:results-so2} and \ce{H_2O} in Section \ref{sec:results-h2o}. The results of incorporating cloud chemistry are presented in Section \ref{sec:results-clouds}.

\subsection{Sulfur-Poor Venus}
\label{sec:results-so2}

\begin{figure*}
\gridline{\fig{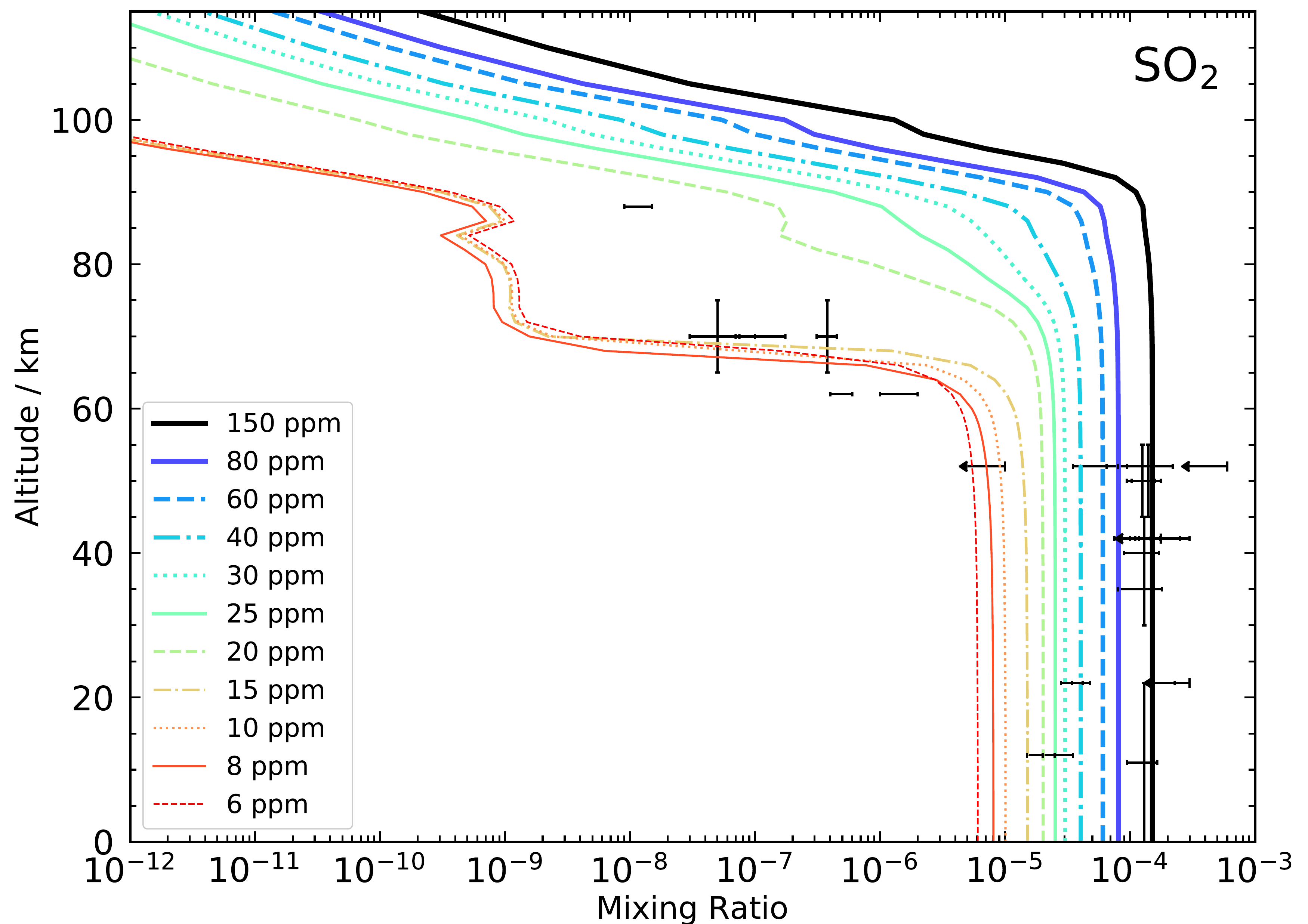}{0.45\textwidth}{(a)}
          \fig{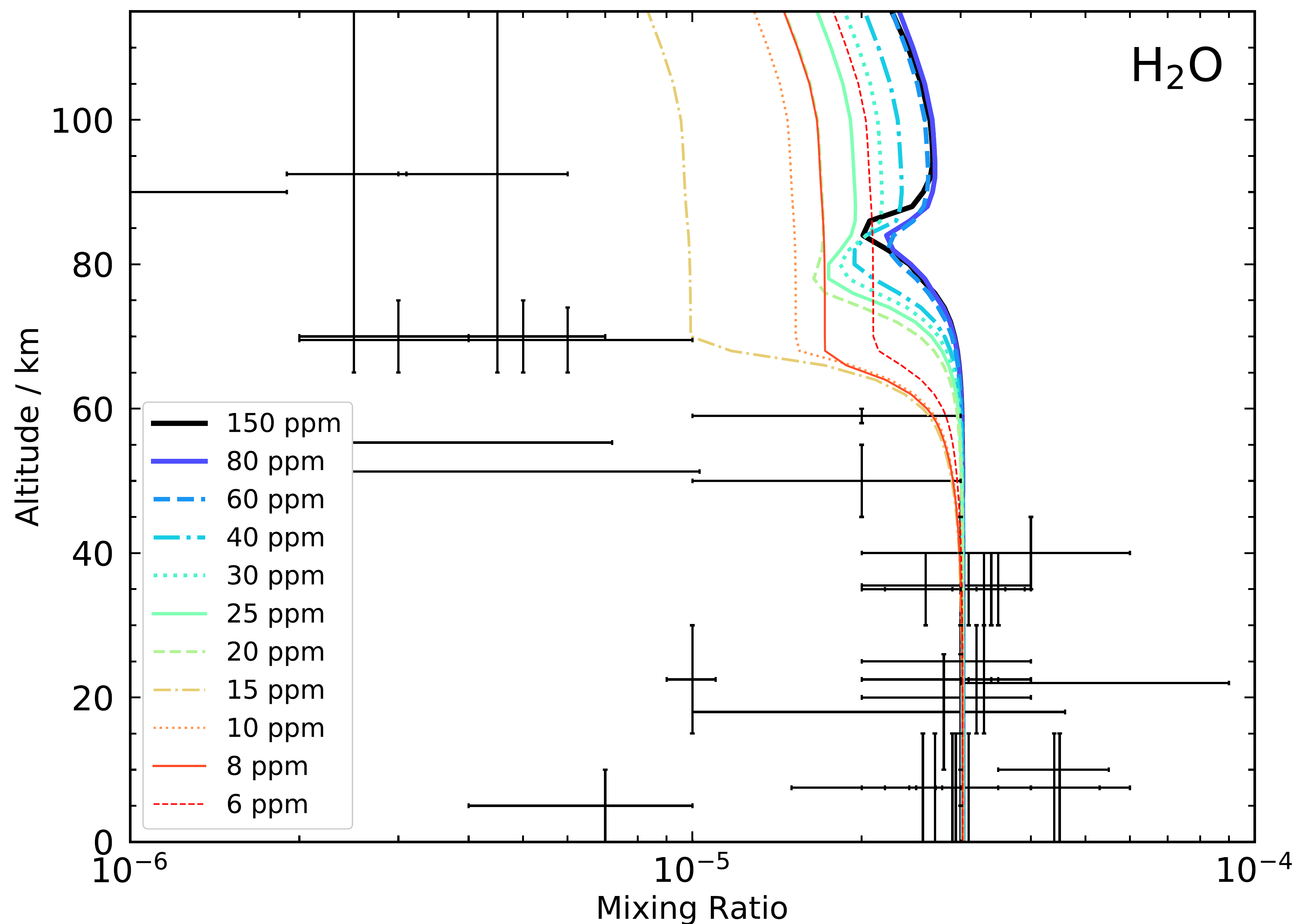}{0.45\textwidth}{(b)}
          }
\gridline{\fig{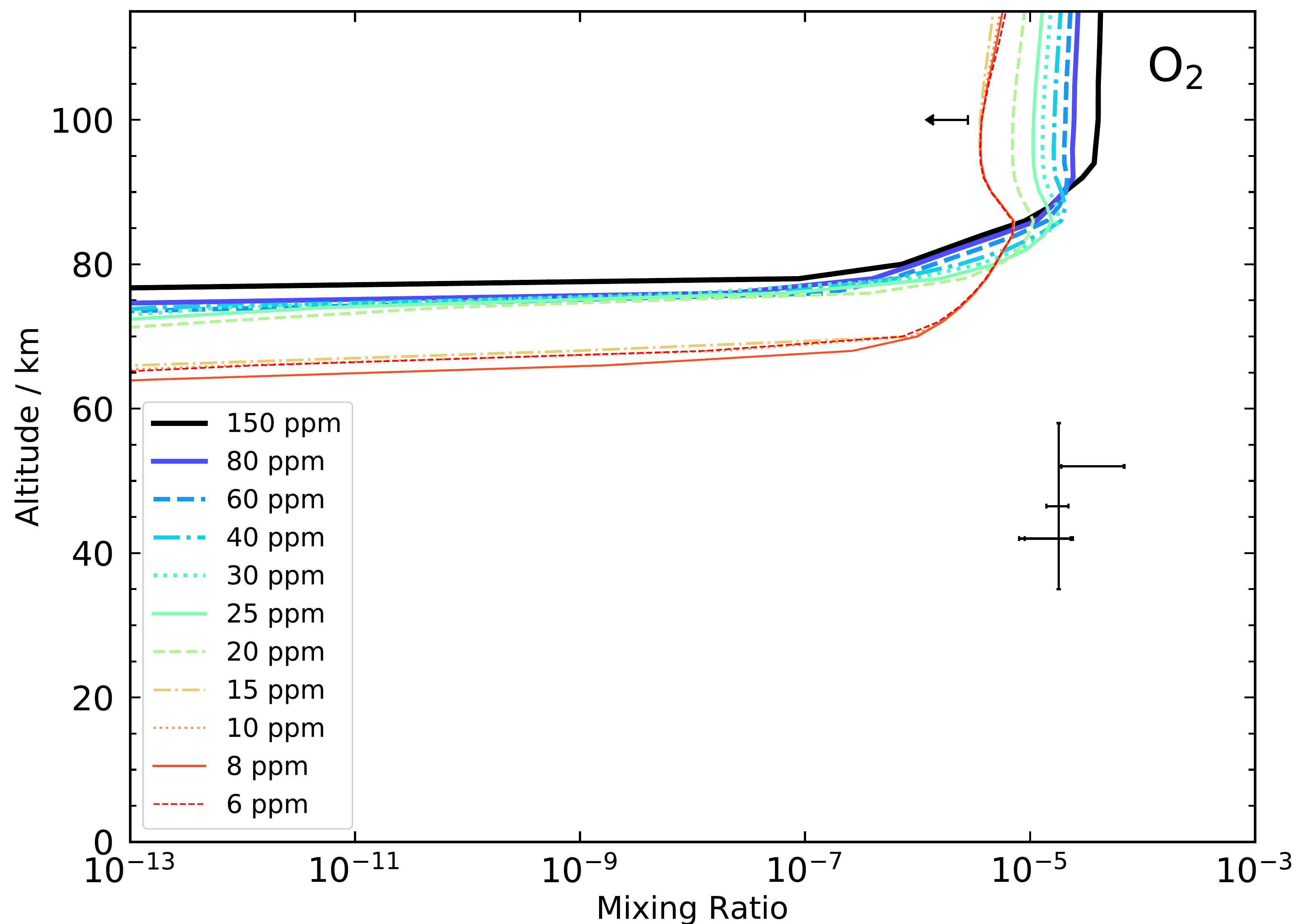}{0.45\textwidth}{(c)}
          \fig{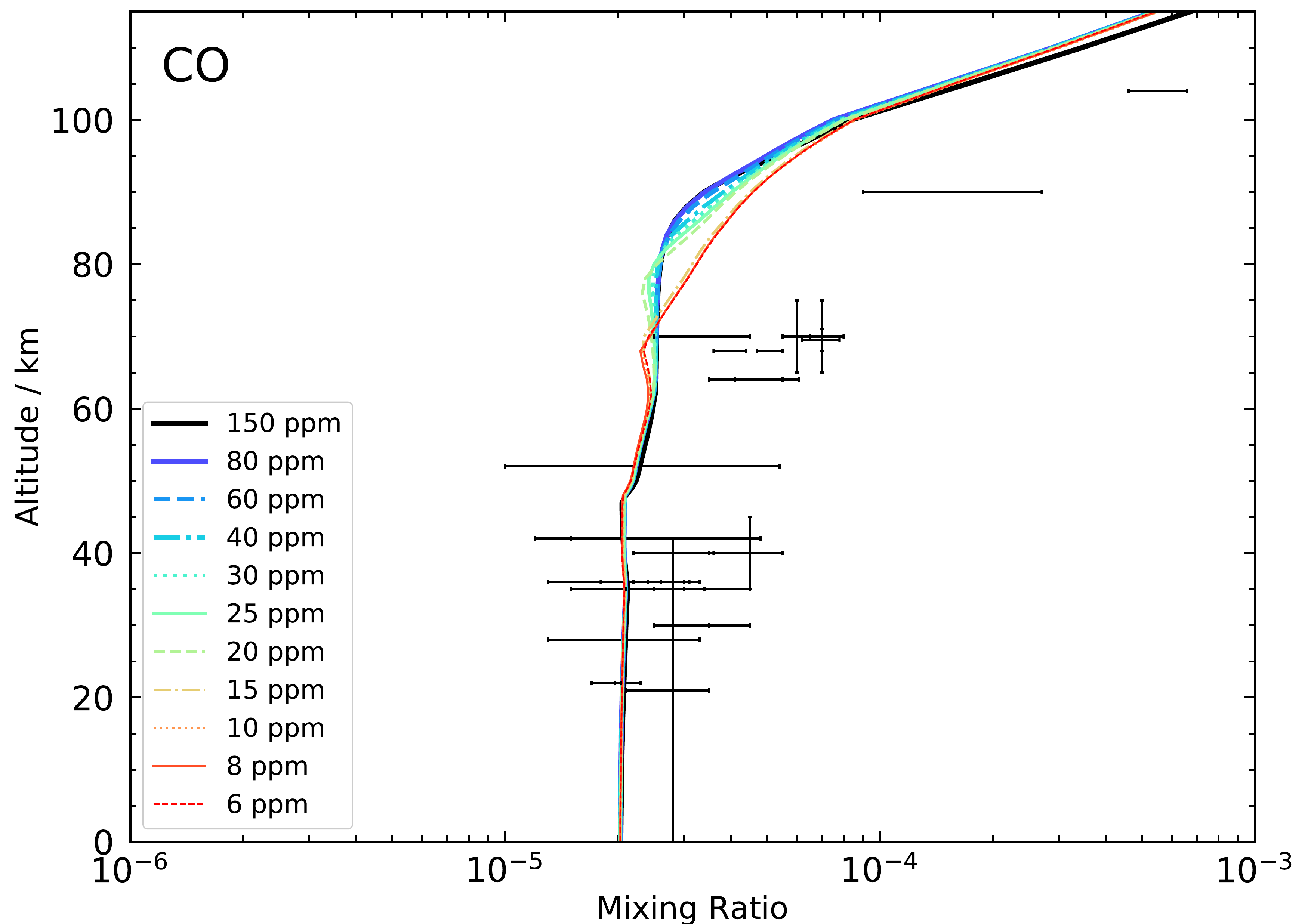}{0.45\textwidth}{(d)}
          }
\caption{Mixing ratios as a function of height for \ce{SO_2} (a), \ce{H_2O} (b), \ce{O_2} (c) and \ce{CO} (d) by varying the below-cloud abundance of \ce{SO_2} from 6 ppm to 80 ppm. No droplet chemistry is included.  \label{fig:so2-poor}}
\end{figure*}

\ce{SO_2} begins to significantly decrease through and above the clouds when surface $f_{\ce{SO_2}} \sim 50 {\rm \, ppm}$, and achieves best results for below-cloud concentrations of 15 ppm. See Figure \ref{fig:so2-poor}. The depletion occurs at 70-80 km unless the surface \ce{SO_2} is lower, around 15 ppm. Besides being inconsistent with most below-cloud observational constraints on $f_{\ce{SO_2}}$ by almost an order of magnitude, the sulfur-poor model predictions agree with observations for all the species we consider reasonably well. The results agree with observations as closely as when we consider cloud droplet chemistry below with one possible exception of \ce{SO}, which is below 1 ppb at between 80 and 100 km. This may be brought into better agreement by including a mesospheric source of sulfur acid vapor, or by adjusting the below-cloud concentrations of \ce{SO_2}, since we have found that the above-cloud \ce{SO_2} is very sensitive to the below-cloud \ce{SO_2} and the Eddy diffusion profile when $f_{\ce{SO_2}} \approx $ 15 ppm. The 15 ppm below-cloud \ce{SO_2} model also predicts \ce{H_2} concentrations of $\sim 10 \, {\rm ppm}$ above 70 km.

\subsection{Water-Rich Venus}
\label{sec:results-h2o}

\begin{figure*}
\gridline{\fig{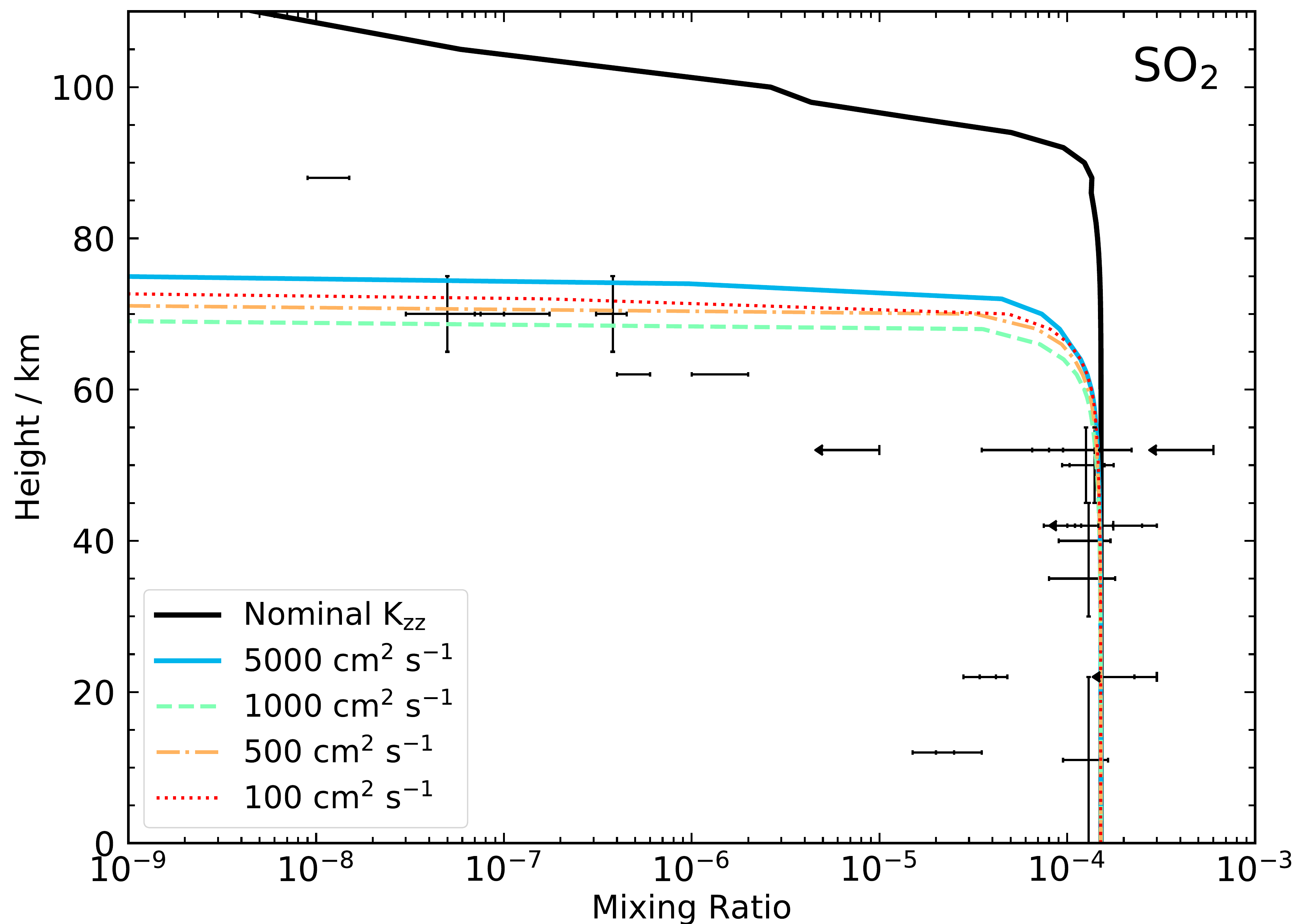}{0.45\textwidth}{(a)}
          \fig{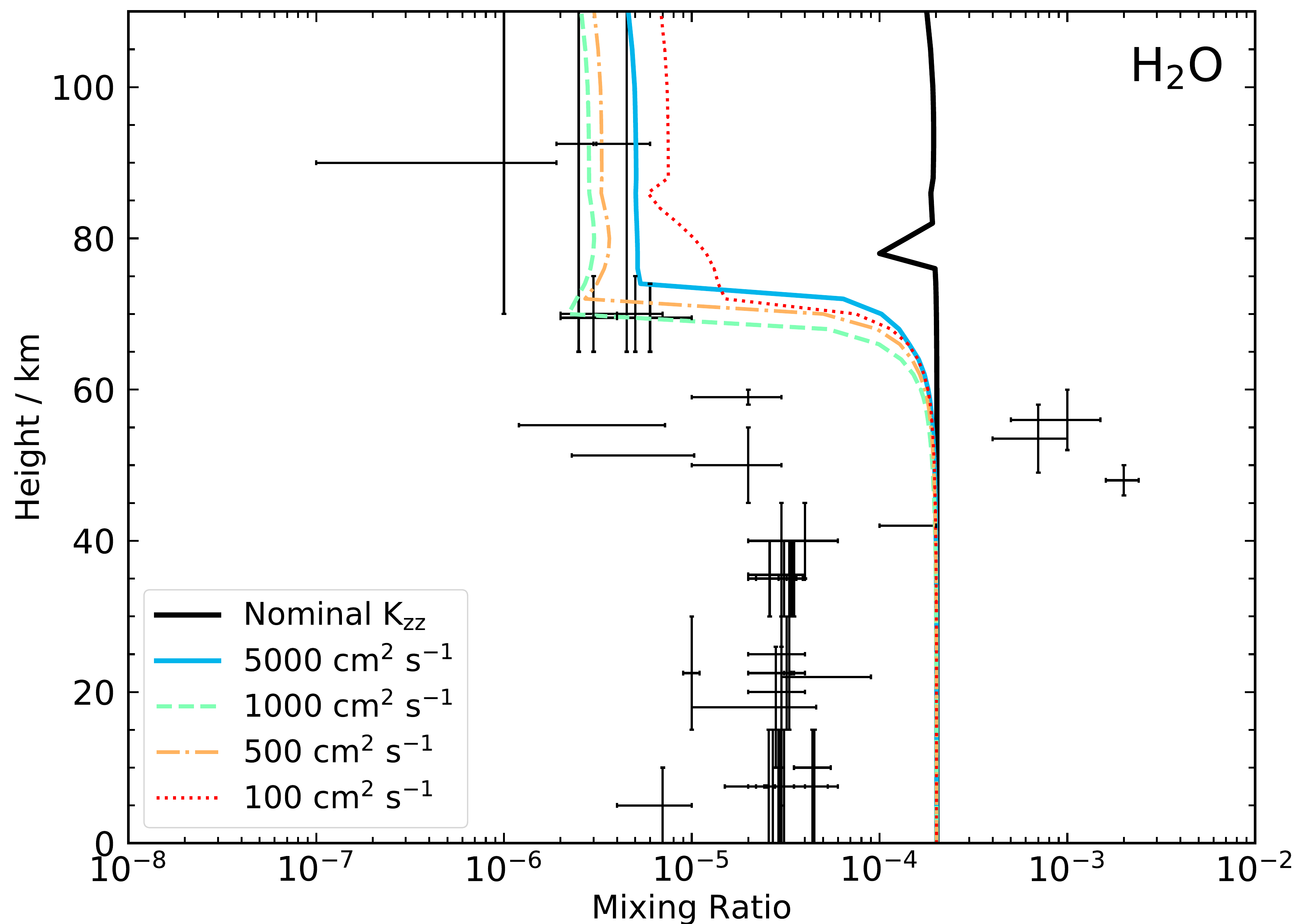}{0.45\textwidth}{(b)}
          }
\gridline{\fig{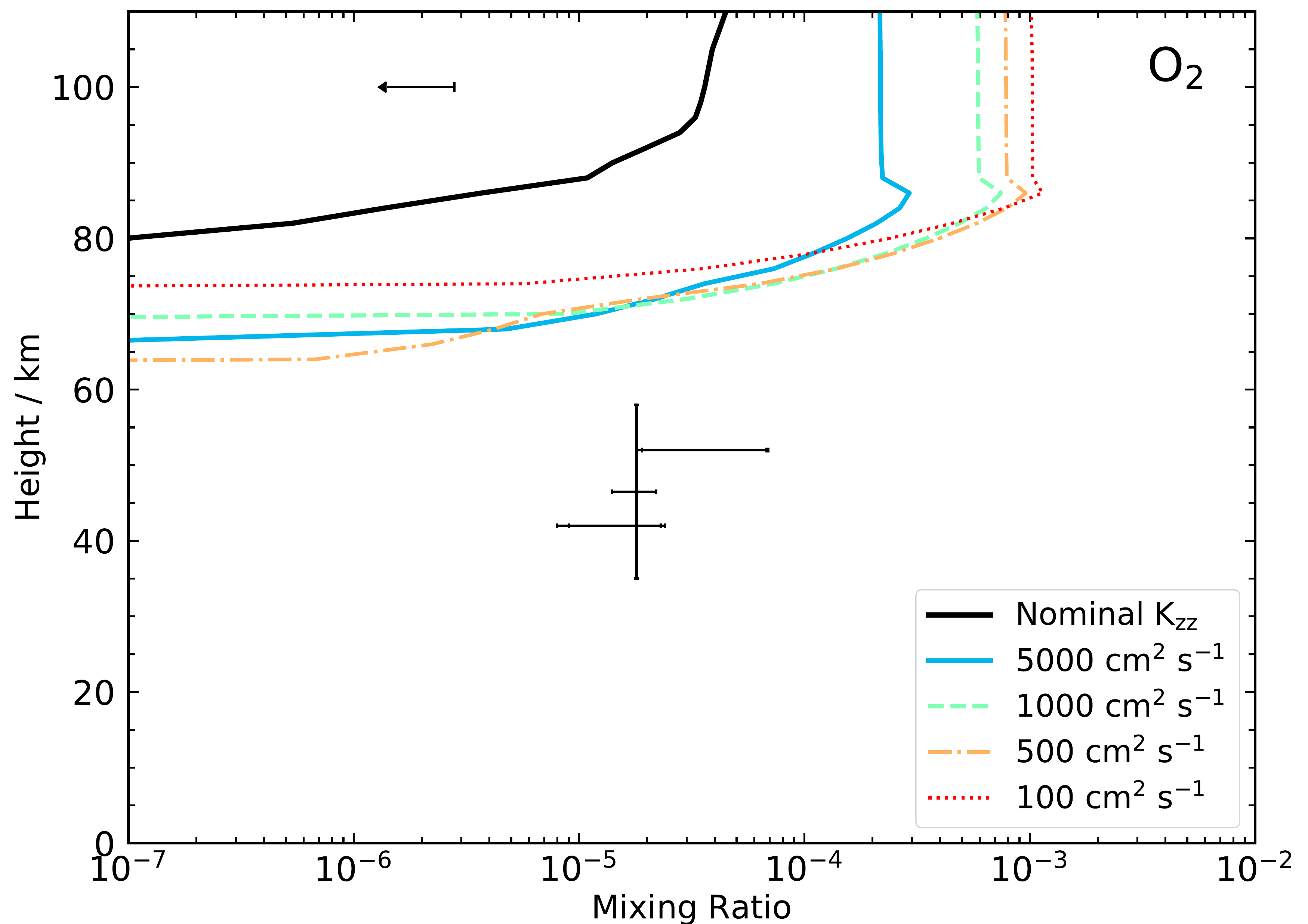}{0.45\textwidth}{(c)}
          \fig{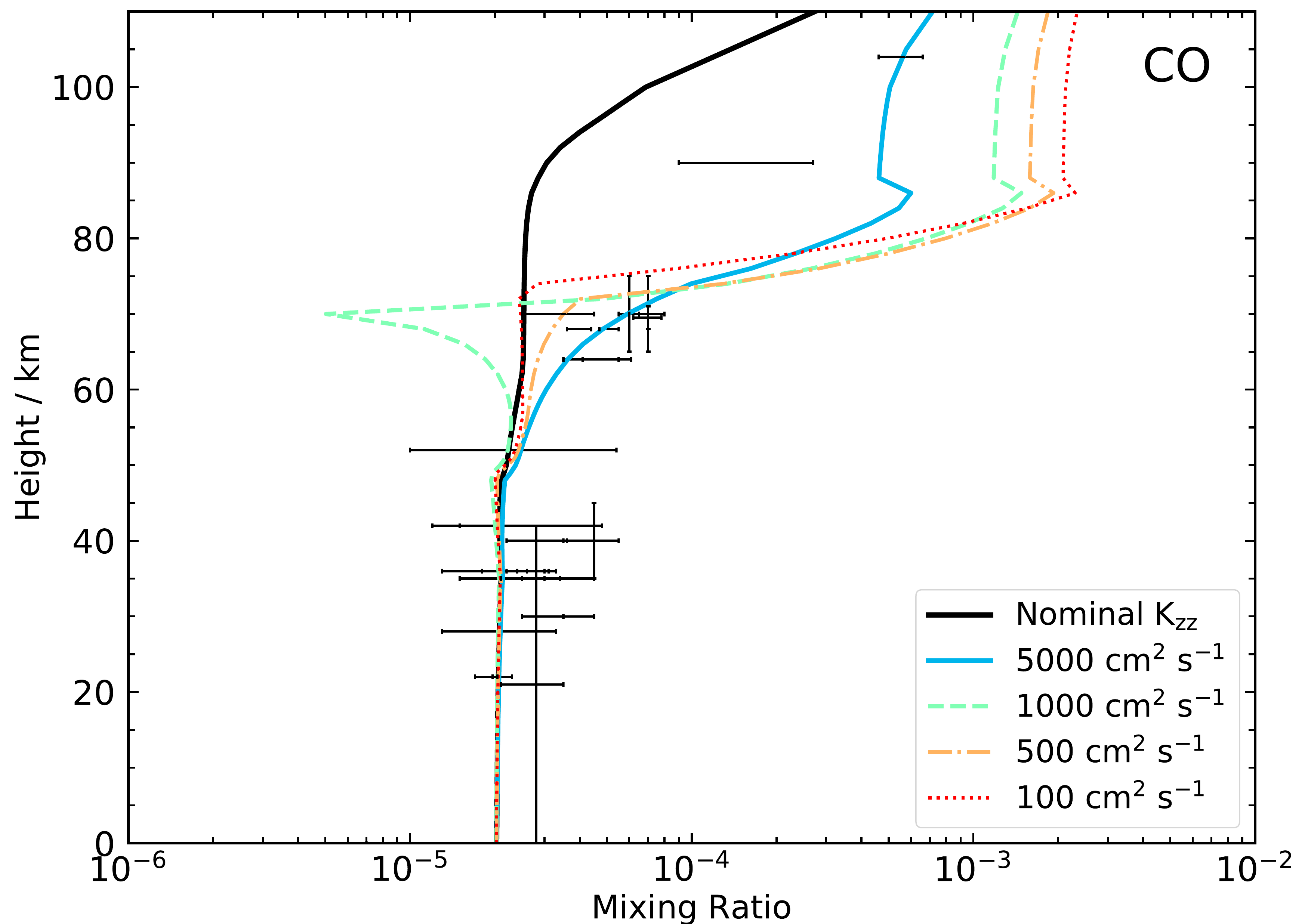}{0.45\textwidth}{(d)}
          }
\caption{Mixing ratios as a function of height for \ce{SO_2} (a), \ce{H_2O} (b), \ce{O_2} (c) and \ce{CO} (d) by varying $K_{zz}$ (see Figure \ref{fig:profiles}) with the abundance of below-cloud \ce{H_2O} of 200 ppm. The divot in the water abundance for nominal \ce{K_{zz}} is due to condensation. The \ce{CO} profile when $K_{zz} = 1000 {\rm \, cm^{2} s^{-1}}$ bears a remarkable similarity to the \ce{CO} profile from \citet{Marcq2005}. No droplet chemistry is included. \label{fig:h2o-rich}}
\end{figure*}

For the water-rich case, \ce{SO_2} does not deplete for any value of below-cloud \ce{H_2O} unless a $K_{zz}$ trap extending to 85 km in altitude is introduced in the eddy diffusion profile. Upon introduction of this trap the \ce{SO_2} begins to significantly deplete at the top of the atmosphere, and this depletion height then lowers with increasing below-cloud water abundance, achieving best results around surface $f_{\ce{H_2O}} = {\rm 200 \, ppm}$. See Figure \ref{fig:h2o-rich}. For in-cloud $K_{zz} = {\rm 5000 \, cm^2 \, s^{-1}}$ the depletion height is higher than observations suggest \citep{Encrenaz2019}, dropping off at $\gtrsim 75{\rm \, km}$, and the \ce{O_2} concentration in the upper atmosphere exceeds 100 ppm. The depletion height can be lowered further to $\sim {\rm \, 70 \, km}$ by decreasing the $K_{zz}$ value in the trap, however this exacerbates the overabundance of \ce{O_2} and causes \ce{CO} to deviate from smooth monotonic growth with altitude. The \ce{CO} profile has a downward spike at $\sim 70 {\rm \, km}$, consistent the profile of \citet{Pollack1993}, as scaled by \citet{Marcq2005}. This is likely due to the change of $K_{zz}$, see Section \ref{sec:discussion}. Neither the \ce{SO_2} depletion or \ce{O_2} abundances agree with observations as well as in the sulfur-poor and cloud-chemistry cases. The reason for this is the self-shielding of the excess water, which is not as effectively removed as \ce{SO_2} within the clouds. The 200 ppm below-cloud \ce{H_2O} model also predicts \ce{H_2} concentrations of $\sim 100 \, {\rm ppm}$ above 70 km, inconsistent with observations (see Appendix \ref{app:obs}).

\subsection{Cloud Chemistry}
\label{sec:results-clouds}

\begin{figure}[ht!]
\plotone{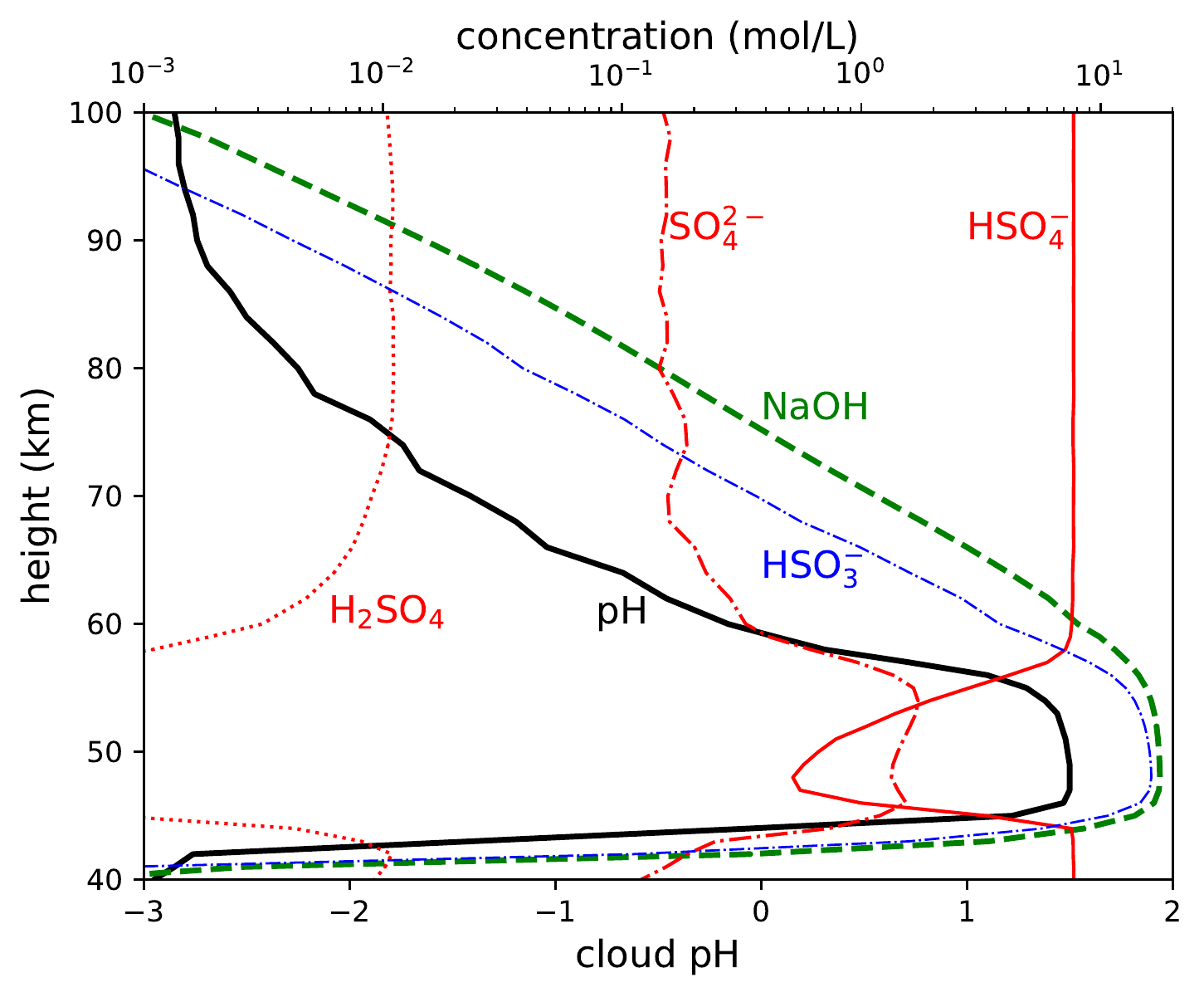}
\caption{Predicted cloud droplet pH (bottom axis, solid black line) initial \ce{NaOH} concentrations (mol/L, top axis, dashed green line), and the speciation of the sulfuric and sulfurous acid (red and blue lines), as a function of height (km) based on \ce{SO_2} depletion. The initial \ce{NaOH} is an input that we use to reproduce the \ce{SO_2} depletion. This solution for \ce{SO_2} is not unique. Changing the pH by changing the initial \ce{NaOH} will affect the \ce{SO_2} depletion. Alternatively, the pH could be higher than plotted, and the depletion could be limited by kinetics. The profiles with \ce{Ca(OH)_2} are similar. \label{fig:cloud-ph}}
\end{figure}

\begin{figure}[ht!]
\plotone{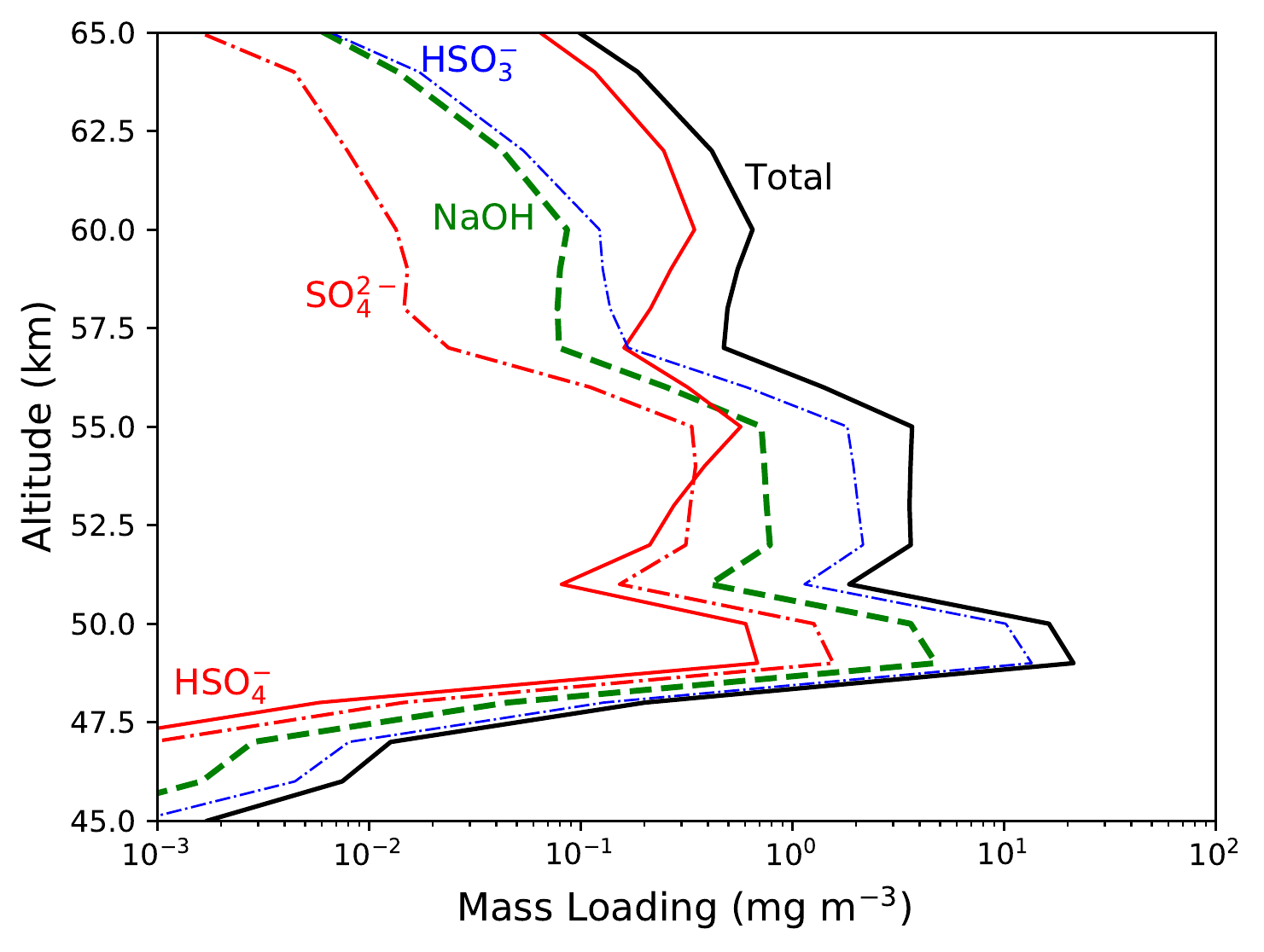}
\caption{Mass composition of the droplets (mass loading, mg m$^{-3}$) as a function of altitude (km). The total mass loading is based on Pioneer Venus observations summed over the three aerosol modes \citep{Knollenberg1980,Kras2016}. The species mass loading is equal to the mass fraction multiplied by the total mass loading. The profiles with \ce{Ca(OH)_2} are similar. \label{fig:massload}}
\end{figure}

In this scenario, described in Section \ref{sec:cloud}, \ce{SO_2} depletion through the clouds is accomplished by removing the \ce{SO_2} into \ce{H_2SO_3} and \ce{H_2SO_4} via droplet chemistry. In our model, the droplet chemistry is driven specifically by sodium hydroxide (\ce{NaOH}), and the \ce{NaOH} itself buffers the cloud pH. The \ce{NaOH} is a proxy used for convenience in modelling, and represents other plausible sources of delivered hydrogen, discussed in Section \ref{sec:discussion}. We adjust the amount of \ce{NaOH} as a function of height to reproduce the observationally constrained \ce{SO_2} profile (see Figure \ref{fig:cloud-ph}). This initial \ce{NaOH} is prescribed for each height and is not solved for within the model. The function that reproduces the \ce{SO_2} depletion, given the estimated droplet volume, was determined by solving the aqueous chemistry and cloud chemistry for different amounts of \ce{NaOH}, and the concentration of \ce{NaOH} and the speciation of sulfuric and sulfurous acid as a function of height that results in the observed \ce{SO_2} profile is shown in Figure \ref{fig:cloud-ph}. The prediction that the bulk of the clouds is \ce{HSO_3^-} only holds if there is no mechanism for oxidizing the sulfur in the clouds. Sulfite aerosols are rapidly oxidized in Earth's atmosphere \citep{Townsend2012}, and this may be the case for the atmosphere of Venus as well. We show the predicted mass loading in Figure \ref{fig:massload}. The mass of the lower cloud is largely in the form of sulfite, and this region overlaps with the larger mode 3 aerosols \citep{Knollenberg1980}. Profiles are similar for \ce{Ca(OH)_2}, assuming the kinetics are identical between \ce{Ca(OH)_2} and \ce{NaOH}. The similarity is due to the large $pK_a$ which results in all the \ce{Ca(OH)_2} becoming fully dissociated, along with the similar mass to basicity potential: the mass of a calcium atom is almost double that of a sodium atom, and each calcium carries two hydrides, which replace two protons. 

We can then consider our solution of the droplet chemistry, set out in Equations (\ref{eqn:liquid-first}) -- (\ref{eqn:liquid-last}), which predicts the \ce{H^+} activity, from which we can calculate the cloud droplet pH. As described in Section \ref{sec:cloud}, the \ce{SO_2} depletion is set by the capacity of the liquid to hold \ce{SO_2}, which is controlled by the amount of \ce{NaOH}, and by the total volume of the liquid, which is determined by the cloud droplet size distribution. The predicted droplet pH is plotted as a function of height in Figure \ref{fig:cloud-ph}.

In this model, throughout the clouds the gas-phase \ce{SO_2} is in equilibrium with the concentration of sulfur in the droplet that is specifically in the form of \ce{SO_2}; i.e., $\mathrm{SO_2(g)}\propto\mathrm{SO_2}(\ell)$ and adding or removing gas-phase \ce{SO_2} results in proportionally changing the droplet \ce{SO_2}, and the balance of the other sulfur species. This equilibrium allows us to write out effective rate constants for \ce{SO_2} and \ce{H_2O} dissolution into the droplets with rate constants tuned to reproduce the results from solving Eq's (\ref{eqn:liquid-first}) -- (\ref{eqn:liquid-last}). For \ce{SO_2} the effective reaction is:
\begin{equation}
\ce{SO_2(g)} + \ce{H_2SO_4($\ell$)} \rightleftharpoons \ce{SO_2($\ell$)} + \ce{H_2SO_4($\ell$)}.
\end{equation}
The rate constants for the forward reaction, $k_f$ (cm$^3$ s$^{-1}$), and reverse reaction $k_r$ (cm$^3$ s$^{-1}$) are:
\begin{align}
k_f &= 10^{-32} \; \mathrm{cm^3 \, s^{-1}} \; e^{9000 \, {\rm K}/T}, \\
k_r &= 6.67 \times 10^{-36} \; \mathrm{cm^3 \, s^{-1}} \; e^{9000 \, {\rm K}/T}.
\end{align}
For \ce{H_2O} the effective reaction is:
\begin{equation}
\ce{H_2O(g)} + 2\ce{SO_2($\ell$)} \rightleftharpoons \ce{H_2SO_3($\ell$)} + \ce{SO_2($\ell$)}.
\end{equation}
Neither of these reactions account for condensation. We determine the effective rate constants for the forward reaction, $k_f$ (cm$^3$ s$^{-1}$), and reverse reaction $k_r$ (cm$^3$ s$^{-1}$) to be:
\begin{align}
k_f &= 2.53 \times 10^{-36} \; \mathrm{cm^3 \, s^{-1}} \; e^{9000 \, {\rm K}/T}, \\
k_r &= 8.43 \times 10^{-38} \; \mathrm{cm^3 \, s^{-1}} \; e^{9000 \, {\rm K}/T}.
\end{align}
Finally, an effective reaction needs to be included to encapsulate the release of \ce{SO_2} when the droplet rains out and evaporates:
\begin{equation}
\ce{SO_2($\ell$)} + \ce{H_2SO_4(g)} \rightarrow \ce{SO_2(g)} + \ce{H_2SO_4(g)},
\end{equation}
with rate constant of $2.2 \times 10^{-4}$ cm$^3$ s$^{-1}$ $e^{-10000 \, {\rm K}/T}$.

\begin{figure}[ht!]
\includegraphics[width=\textwidth]{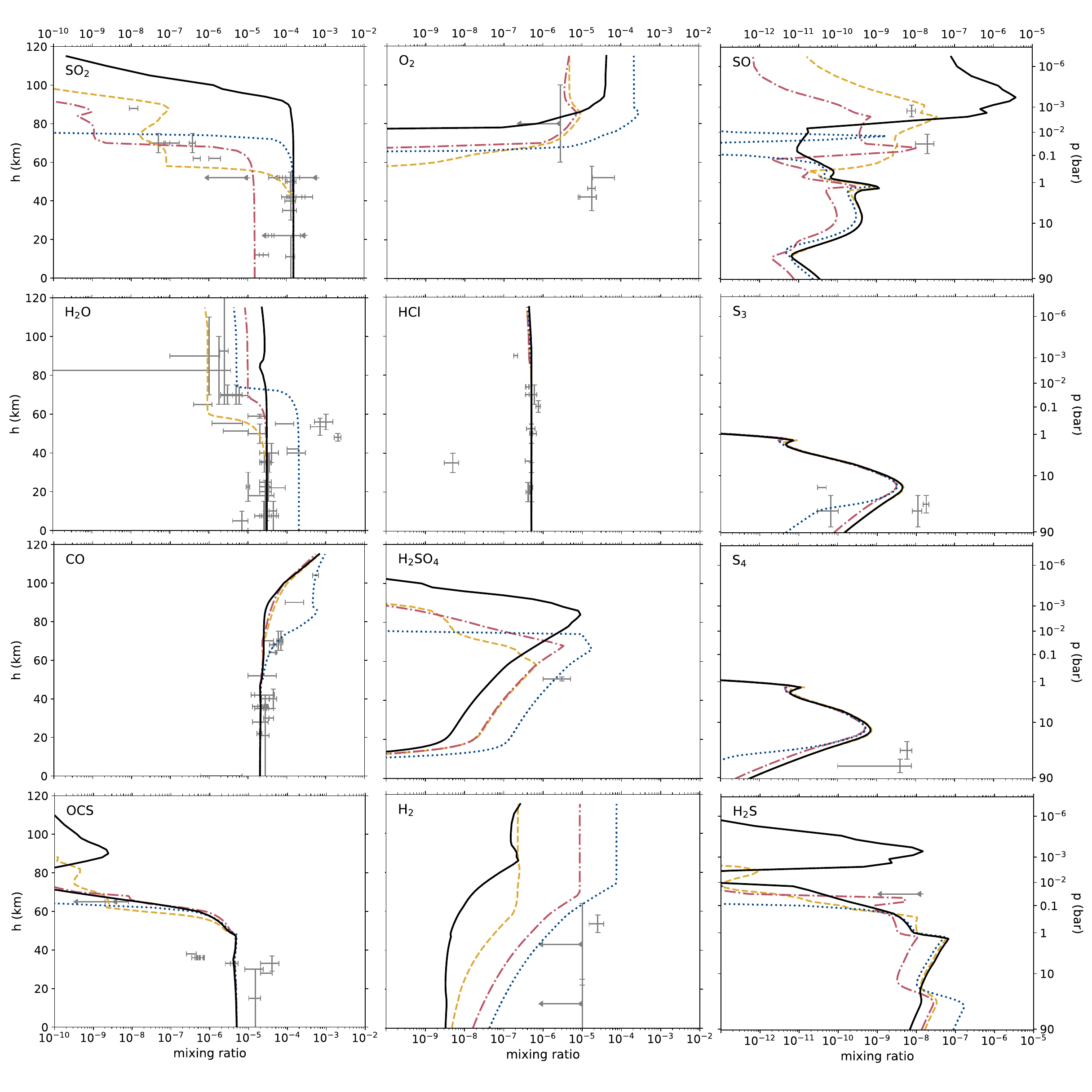}
\caption{Predicted volume mixing ratios of \ce{SO_2}, \ce{H_2O}, \ce{CO}, \ce{OCS}, \ce{O_2}, \ce{HCl}, \ce{H_2SO_4}, \ce{H_2}, \ce{SO}, \ce{S_3}, \ce{S_4} and \ce{H_2S} as a function of height (km), and compared to data (see Appendix \ref{app:obs}), for four models: the cloud-chemistry model (yellow dashed, Sections \ref{sec:cloud} and \ref{sec:results-clouds}), the sulfur-poor model with $f(\ce{SO_2}) = 20\,{\rm ppm}$ (red dash-dot, Section \ref{sec:results-so2}), the water-rich model with $f(\ce{H_2O}) = 200\,{\rm ppm}$ and in-cloud $K_{zz} = 5000\,{\rm cm^{2}\,s^{-1}}$ (blue dotted, Section \ref{sec:results-h2o}), and the fiducial model with $f(\ce{SO_2}) = 150\,{\rm ppm}$, $f(\ce{H_2O}) = 30\,{\rm ppm}$ and no cloud chemistry (black solid). Observations and upper limits of these species (from Table \ref{tab:obs}) are also plotted in gray to compare. \ce{H_2SO_4} includes both condensed and gas-phase \ce{H_2SO_4}. \label{fig:cloudchem}}
\end{figure}

The results of our model agree within an order of magnitude for all species considered, and within a factor of 3 for all species except for \ce{OCS} and the sulfur allotropes. Our model underestimates \ce{S_4} and does not predict the steep below-cloud gradient of \ce{OCS}. In addition, it over-predicts \ce{O_2} in the upper atmosphere by a factor of 2-3. Comparison of this model and the best sulfur-poor model, with $f(\ce{SO_2}) = 20$ ppm and Krasnopolsky's Eddy diffusion profile (from Section \ref{sec:results-so2}) is shown in Figure \ref{fig:cloudchem}.

\section{Discussion} 
\label{sec:discussion}

Below-cloud \ce{SO_2} $<50\,\mathrm{ppm}$ is inconsistent with most observations but not all. Vega 1 and 2 observed $\ge100\,\mathrm{ppm}$ concentrations of \ce{SO_2} directly below the clouds, though the error bars are large and $50\,\mathrm{ppm}$ abundances would be within $2\sigma$ of the measurements \citep{Bertaux1996}. Even lower \ce{SO_2} was measured within 20 km of the surface \citep{Bertaux1996}, which could indicate rapid surface depletion of \ce{SO_2} \citep{Yung1982}.

The reported observations of below-cloud \ce{SO2} at 100--$200\,\mathrm{ppm}$ also have large uncertainties, typically on the order of 50 ppm \citep[see, e.g.][]{Marcq2008}, so true values below 50 ppm would amount to discrepancies of 2 -- 3 $\sigma$. Exploring the hypothesis of low below-cloud \ce{SO_2} as an explanation for the above-cloud \ce{SO_2} depletion will require both more precise and more frequent observations of the below-cloud \ce{SO_2}, to see whether it varies and by how much.  Such data may only be obtainable with {\it in situ} measurements.

Below-cloud \ce{H_2O} is better constrained, so it is less likely that there is an undetected large source of water vapor beneath the clouds. There are some variations in the measurements, from less than 10 ppm to 60 ppm, but with relatively small error bars. There is some observational support for a large reservoir of water within the clouds, with 700 -- 2000 ppm concentrations observed in the cloud layers by Vega 1,2 \citep{Surkov1987} and Venera 13,14 \citep{Mukhin1982,Surkov1982}, and 100 -- 200 ppm concentrations observed from the ground \citep{Bell1991}. These high quantities may have been due to incidental sampling of cloud droplets which are expected to be composed of $\sim 15-25 \%$ water, even without considering the cloud pH buffer hypothesis proposed here. Even if these constraints were not so tight, the model where we increase the below-cloud water vapor predicts that the clouds to extend up to $\sim 80$ km, and the depletion would be more gradual and higher in the atmosphere. Further to this, the water-rich model overproduces \ce{O_2} above $\sim 70$ km pushing it beyond the observed upper limit by two orders of magnitude, overproduces \ce{H_2}, and does not allow for \ce{SO_2} to return at the $\sim 1 {\rm \, ppb}$ level at 90 km, problems which are not shared by the other best fitting models.

In addition to enhancing the cloud layers, we had to vary the above-cloud $K_{zz}$ in order to induce sulfur and water vapor depletion. Decreasing the $K_{zz}$ to $\sim 1000 \, {\rm cm^{2} s^{-1}}$ creates a negative spike in the \ce{CO} at $\sim 70 \, {\rm km}$, reproducing a feature seen in suggested mesospheric profiles \citep{Marcq2005}. This is a consequence of varying the $K_{zz}$ and not increasing the below-cloud water vapor. Our model suggests that this \ce{CO} negative spike may trace changes in the eddy diffusion, a hypothesis that is worth further investigation but is outside the scope of this paper.

Both the sulfur-poor and water-rich models predict $> 10 \, {\rm ppm}$ \ce{H_2} concentrations in the mesosphere, above 70 km, where the cloud chemistry model predicts $\sim 0.1 \, {\rm ppm}$ \ce{H_2} concentrations. Better observational constraints of \ce{H_2} may be useful for distinguishing these models.

As we have shown, cloud chemistry is a possible explanation for the depletion of \ce{SO_2}. Aerosols could provide the excess hydrogen capable of depleting gas-phase \ce{SO_2}. This hydrogen could be bound in salts, or could be in some other form, such as hydrocarbons. It is important that whatever the source of hydrogen, it is replenished to sustain the \ce{SO_2} gradient. Otherwise the hydrogen will be consumed, the clouds will be saturated with sulfur, and the gradient of \ce{SO_2} will disappear. It is also worth mentioning that three phases can participate in this chemistry: the gas phase, the liquid of the droplet (at atmospheric heights where the droplet has not frozen), and the solid aerosol material, either at the core, surrounding the droplet, or suspended within the droplet. This provides a rich and complex multi-phase chemistry worth exploring in the lab.

What follows in this section is a discussion of the requirements for and implications of cloud chemistry. In Section \ref{sec:sources}, we discuss possible sources of hydrogen within the clouds. The measured optical constants and spectral features of the clouds of Venus are consistent with cloud droplets composed of a large percentage of sulfuric acid. Any proposed cloud chemistry must either preserve sulfuric acid as the dominant species in the clouds, or must propose a species with similar optical properties and spectral features. We discuss the observable implications of our cloud chemistry in Section \ref{sec:cloud-obs}. The dissolution of hydrogen halides into the clouds is discussed in Section \ref{sec:cloud-hal}. The cloud chemistry also has implications for above-cloud radical concentrations, which affects the lifetime of hypothetical \ce{PH_3} within the clouds. We discuss the status of \ce{PH_3} and its lifetime within the clouds in Section \ref{sec:ph3}. Finally, in Section \ref{sec:implications-life}, we briefly discuss the implications of different cloud chemistry for hypothetical life within the cloud droplets of Venus.

\subsection{Possible Sources of Hydrogen in the Clouds}
\label{sec:sources}

If a buffer explains the sulfur depletion, it is possibly a salt. Salts will dissociate quickly, and some will provide efficient buffers. The salts must get into the clouds in order to buffer them. This can be accomplished either by transport from the surface or exogenous delivery. 
\vspace{5mm}\\
{\bf Exogenous Delivery:} Exogenous delivery, meaning delivery of material from the interplanetary medium, is unlikely to provide significant material to buffer the clouds of Venus based on the estimated incoming flux of interplanetary dust. The clouds must be able to retain virtually all of the \ce{SO_2} over the timescale of transport through the clouds, requiring a flux of salts, $\Phi_s$ (mol cm$^{-2}$ s$^{-1}$), of:
\begin{equation}
\Phi_s = \dfrac{n_{\ce{SO_2}} \, H_0}{\tau_{\rm dyn}} = \dfrac{f_{\ce{SO_2}} \, nkT \, K_{zz}}{\mu_{\rm av} m_p \, N_A \, g \, (\Delta h)^2},
\end{equation}
where $n_{\ce{SO_2}} \, ({\rm cm^{-3}})$ is the number density of \ce{SO_2} at the height where \ce{SO_2} begins to deplete (50 km), $H_0 \, {\rm (km)}$ is the scale height, $R = 6052 \, {\rm km}$ is the radius of Venus, $\tau_{\rm dyn} \, {\rm (s)}$ is the dynamic timescale of the atmosphere at 40 km, $f_{\ce{SO_2}} = 150$ ppm is the volume mixing ratio of \ce{SO_2}, $n = \Sigma_{\ce{X}} n_{\ce{X}} = 2.189 \times 10^{19}$ cm$^{-3}$ is the gas density at 50 km, the height where the depletion begins, $k = 1.38065 \times 10^{-16}$ erg/K is Boltzmann's constant, $T = 349.7$ K is the temperature at 40 km, $K_{zz} \approx 100$ cm$^2$ s$^{-1}$ is the Eddy diffusion of the droplets within the cloud, a low estimate more favorable for exogenous delivery. $\mu_{\rm av} = 44$ is the mean molecular weight of the atmosphere, $m_p = 1.6726 \times 10^{-24}$ g is the mass of a proton, $N_A = 6.022 \times 10^{23}$ is Avogadro's Number, $g = 8.87$ m s$^{-2}$ is the surface gravity of Venus and $\Delta h = 20$ km is the thickness of the cloud layer. Applying all these estimates yields:
\begin{equation}
\Phi_s \approx 10^{-13} \; \mathrm{mol \, cm^{-2} \, s^{-1}}.
\end{equation}
Continuous exogenous delivery of material is insufficient to match this flux even assuming that 100\% of the material is in the form of hydrated minerals that will deplete \ce{SO_2}. If we assume that delivery of exogenous material to Venus is comparable to delivery to Earth, 20 -- 70 ktonnes/year \citep{Peucker1996,Greaves2020}, which translates to $\sim 10^{-17}$ mol cm$^{-2}$ s$^{-1}$, or four orders of magnitude too little to account for the missing hydrogen.

One other possibility is stochastic delivery. If a recent airburst, a large impact that breaks up in the atmosphere, occurred in Venus's atmosphere, the metals released could permeate the clouds, providing a transient in-cloud source of hydrogen. If this is the case, then the depletion of \ce{SO_2} will be temporary, lasting as long as the below-cloud store of these elements persists, on the order of the diffusion timescale or $\sim 1000$ years.
\vspace{5mm}\\
{\bf Dust from the Surface:} Transport from the surface also struggles to meet the required flux of hydrated material. Although calculations of dust transport favor a more dusty atmosphere for Venus than Earth \citep{Sagan1975}, Venera measurements suggest that the lower atmosphere is clear, placing upper limits on dust transport to the clouds. It is possible that there is heterogeneous low atmospheric weather, with dust rapidly transported to tens of km above the surface, and that Venera happened to land in a region where the vertical diffusion and winds were insufficient to change the atmospheric opacity. A low level haze inferred from Venera probe data \citep{Grieger2004}, and consistent with heavy metal frost at higher elevations on Venus's surface \citep{Schaefer2004}, may itself be the suspended dust \citep{Titov2018}, and winds may cause that dust to periodically move into the clouds. The reaction products of the salt with \ce{SO_2} in and below the clouds could result in transient high abundances of water, which may then be removed by the left-over oxides. This could reconcile the transient observations of $>100\,{\rm ppm}$ of water vapor in and below the clouds \citep{Mukhin1982,Surkov1982,Bell1991}, with the $\sim 25\,{\rm ppm}$ abundances near the surface \citep{Bertaux1996}.

To explain the sulfur depletion, with dust containing 5 wt.\% salt, requires a dust flux to the clouds of $\approx 16$ Gt/year, well within the estimates of surface dust fluxes estimated from analogue experiments \citep{Greeley1984}. The composition of the dust and the form of the salts is unknown. Here we will speculate on some possible candidates. Our speculation is based on chemical and physical stability of the salts: 
\begin{itemize}
\item {\bf \ce{NaOH}:} Sodium hydroxide is the example salt we use for our calculations, but is an unrealistic candidate salt. It is sufficiently stable to heat, persisting as a liquid up to 1661 K \citep{Haynes2014}. However, it is known to react rapidly with \ce{SO_2} to form sodium sulfite and water vapor. Given the concentrations of below-cloud \ce{SO_2}, \ce{NaOH} cannot plausibly survive to reach the clouds unless it is injected rapidly, e.g. via a volcanic plume.
\item {\bf \ce{Ca(OH)_2}:} Calcium hydroxide is stable as a solid at the surface of Venus. Strictly speaking, calcium hydroxide has no melting point. Instead, at 93 bar and $\sim 1000$ K it is expected to decompose into \ce{CaO} and \ce{H_2O}, based on extrapolations of the vapor pressure curve of \citet{Halstead1957}. \ce{Ca(OH)_2} will undergo carbonation, and there is ample \ce{CO_2}, but this reaction is very slow, and \ce{Ca(OH)_2} is kinetically stable at temperatures above 723 K \citep{Materic2011}. \ce{Ca(OH)_2} will also react rapidly with \ce{SO_2}, but only in the presence of water vapor at concentrations of $>3000$ ppm \citep{Liu2010}. The kinetic stability vs. the dynamic timescale for \ce{Ca(OH)_2} aerosols is unknown, but presently \ce{Ca(OH)_2} cannot be ruled out as a candidate.
\item {\bf \ce{Mg(OH)_2} and \ce{Fe(OH)_2}:} Neither magnesium hydroxide nor iron hydroxide (either the Fe(II) hydroxide or Fe(III) hydroxides) are stable at Venus's surface temperature and pressure \citep{Wang1998,Haynes2014}.
\item {\bf Other Hydroxides:} It may be that more exotic hydroxides, such as \ce{Al(OH)_3}, could deliver hydrogen to Venus's cloud layer. The requirements are sufficient concentrations to satisfy the required fluxes, and the thermochemical and kinetic stability of the salt in the presence of major atmospheric constituents.
\item {\bf Oxides:} Oxides, either resident surface oxides or oxides produced by the dehydration of hydroxides, may participate in cloud chemistry in unknown ways, sequestering \ce{SO_3} directly, for example, \ce{\{Mg,Fe\}O} + \ce{SO_3} may be converted into \ce{\{Mg,Fe\}SO_4} directly. The subsequent dissociation in sulfuric acid will buffer the cloud droplet pH. There is some indirect evidence of the presence of oxides from the near-surface haze, since these oxides can react with hydrochloric acid vapor to form \ce{FeCl_3}, which has been observed in the clouds and is a candidate for the mysterious UV absorber \citep{Kras2017}.
\item {\bf Sulfates:} It is possible hydrogen-bearing sulfates could find their way into the clouds, but no known hydrogen-bearing sulfate is thermally stable at Venus's surface pressure and temperature. It is possible that they are produced from gas-phase reactions, e.g., the possible production of ammonium sulfate from reaction with \ce{NH_3} and \ce{SO_2} \citep{Titov1983}. Sulfates that do not contain hydrogen are plausible aerosols. Indeed, we predict they are produced via cloud chemistry. But these aerosols will not deliver hydrogen and cannot directly participate in the depletion of gas-phase \ce{SO_2}.
\end{itemize}
The above list is not intended to be exhaustive, and does not consider whether these salts are expected at the temperatures and pressures at the surface of Venus, where we would generally expect chemistry to tend toward thermodynamic equilibrium. The results of comparing our surface boundary conditions (Section \ref{sec:model-initial}) to chemical equilibrium (Section \ref{sec:equilib}) predicts that no phyllosilicates or hydrated minerals are present at the surface of Venus if the surface and gas are in equilibrium.

\begin{figure}
  \begin{tabular}{cc}
  \hspace*{-10mm}  
  \includegraphics[width=90mm,trim=200 140 200 130,clip]{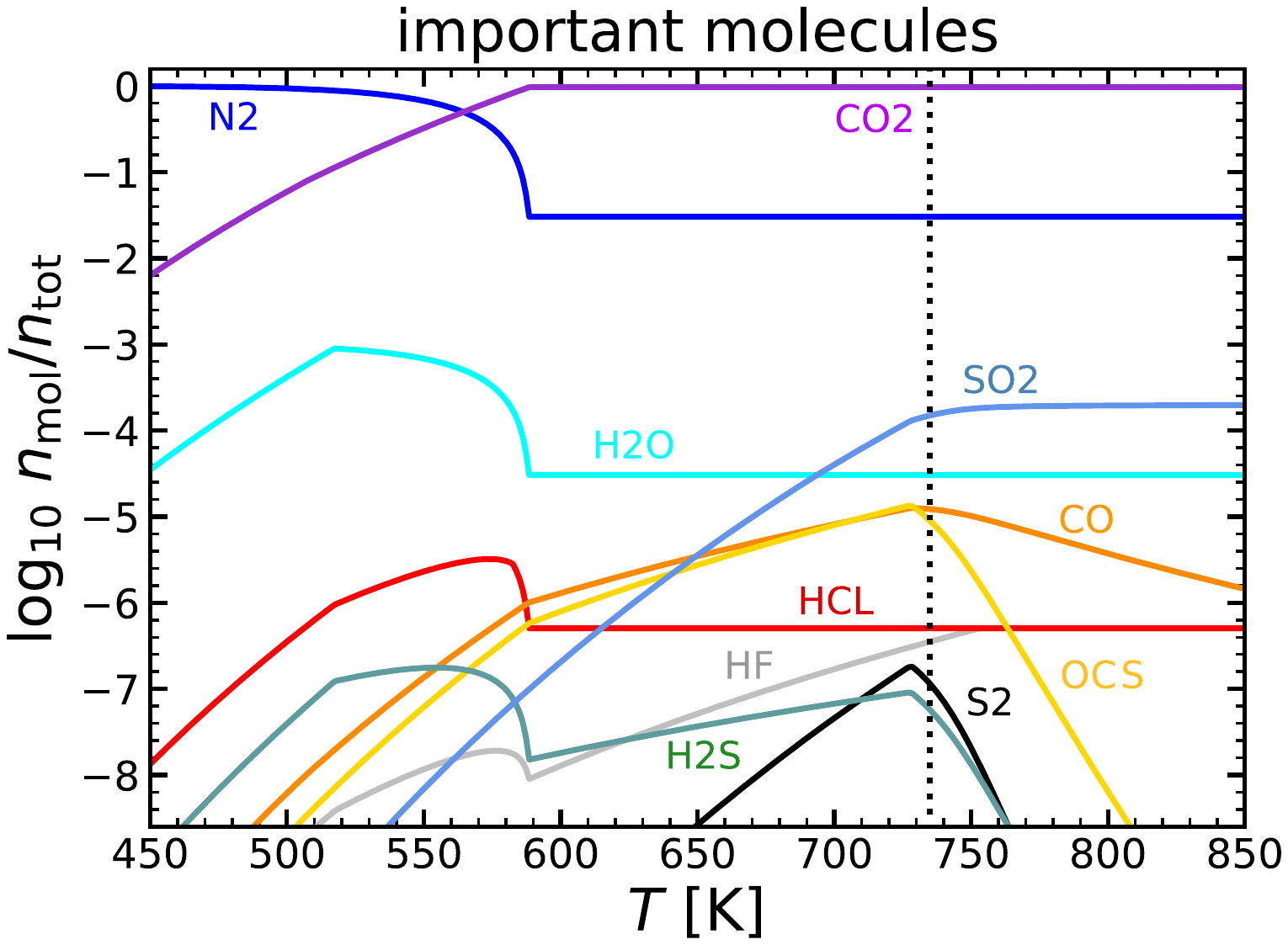}
  \hspace*{-5mm} 
& \includegraphics[width=90mm,trim=200 140 200 130,clip]{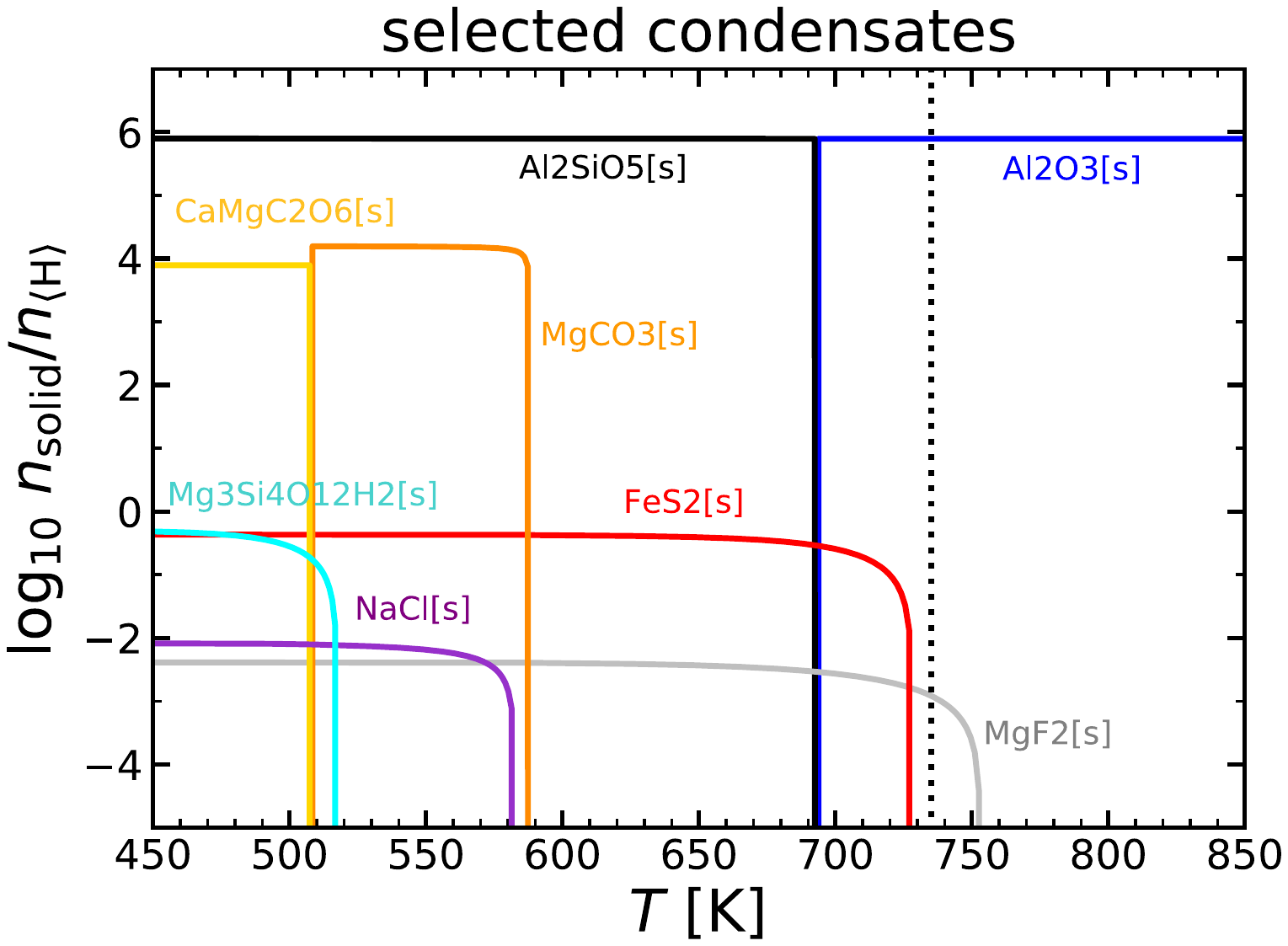}
  \end{tabular}  
  \caption{{\sc GGchem} model for the bottom of the Venus atmosphere, assuming chemical and phase equilibrium, as a function of surface temperature. The temperature of the current Venus surface (735\,K) is marked as dotted vertical lines. A constant gas pressure of $90$\,bar and constant total element abundances $\epsilon^0$ are assumed. $n_{\rm mol}/n_{\rm tot}$ are the molecular concentrations in the gas, and $n_{\rm solid}/n_{\langle\rm H\rangle}$ are the densities of solid species per hydrogen nuclei density. Only solid species which change substantially with temperature are shown. Other stable solids are \ce{MgSiO3[s]}, \ce{SiO2[s]}, \ce{CaAl2Si2O8[s]}, \ce{NaAlSi3O8[s]}, \ce{KAlSi3O8[s]}, \ce{CaSO4[s]}, \ce{Fe2O3[s]}, \ce{TiO2[s]} and \ce{Mn3Al2Si3O12[s]}, which have about constant concentrations as listed in Table~\ref{tab:GGchem}.}
  \label{fig:GGchem}
\end{figure}

The model presented in Sect.~\ref{sec:equilib} can be used to explore the sensitivity of the mineral composition to surface temperature. Figure~\ref{fig:GGchem} shows the results of our {\sc GGchem} model  at the same constant pressure and total element abundances when varying the surface temperature between 450\,K and 850\,K.  Venus is just about 15\,K too warm to have pyrite (\ce{FeS2[s]}) as a stable condensate on the surface according to this model. For surface temperatures lower than about 720\,K, the formation of \ce{FeS2[s]} would start to remove the \ce{SO2} from the atmosphere according to the following complex net reaction
\begin{equation}
\frac{15}{11}\,\ce{SO2}
\,+\, \frac{1}{11}\,\ce{Fe2O3[s]}
\,+\, \ce{CaAl2Si2O8[s]} 
~\longrightarrow~
\frac{2}{11}\,\ce{FeS2[s]}
\,+\, \ce{CaSO4[s]}
\,+\, \ce{Al2O3[s]}
\,+\, 2\,\ce{SiO2[s]} \ ,
\end{equation}
which is a variant of Eq.\,(\ref{eq:regulation}) where the oxygen 
on the left side is provided by \ce{Fe2O3[s]}. At temperatures below about 580\,K in this model, the first carbonate magnesite (\ce{MgCO3[s]}) becomes stable, which could initiate a dramatic change of the atmosphere as the main atmospheric molecule \ce{CO2} could deposit at the surface to form \ce{MgCO3[s]}:
\begin{equation}
\ce{CO2} \,+\, \ce{MgSiO3[s]} ~\longrightarrow~ 
\ce{MgCO3[s]} \,+\, \ce{SiO2[s]} \ ,
\end{equation}
leaving an atmosphere that is dominated by \ce{N2} with more \ce{H2O}. Finally, below a temperature of about 520\,K, the first phyllosilicate, talc (\ce{Mg3Si4O12H2[s]}) becomes stable, which could partly remove the water from the atmosphere. Only for temperatures below 520\,K our model predicts the presence of hydrogen-containing minerals.

Whether or not this means that the gas at the bottom of the Venusian atmosphere is in fact in chemical equilibrium, and whether the element abundances in the gas are regulated by outgassing/deposition via the contact with the hot rock at the surface are yet unsolved questions. Disequilibrium processes might supply phyllosilicates or hydrated salts. For example, geological processes such as volcanism may resurface Venus's crust with hydrated components (see also discussion below). Our phase-equilibrium model suggests that hydrogenated rock and salts are not stable on the surface of  Venus and will sublimate or react with the atmospheric gases to form other chemicals.  However, the salts may be stable enough to be swept up to greater altitudes and cooler temperatures, before they react away and equilibrium is restored.

More research in the lab and {\it in situ} observations of the clouds of Venus will be needed to determine if salts are present and, if so, what their chemistry is.
\vspace{5mm}\\
{\bf Volcanic Delivery:} A variant of the dust delivery mechanism to achieve a hydrogen flux to the clouds is volcanism.  The presence of active volcanism on Venus has long been speculated upon, motivated both by the transient \ce{SO2} detected at 40 mbar by Pioneer Venus \citep[Table \ref{tab:obs};][]{esposito1984_science} and the recognition that \ce{SO2} may react with surface minerals and require continual replenishment \citep{fegley1989_nature}.

Volcanism could deliver material to Venus's clouds in three ways: 1) as solid material deposited at the surface, which is subsequently lofted by winds; 2) as an explosive eruption introducing material into the below-cloud atmosphere, where vertical mixing slowly raises it into the cloud layer; and 3) as a large explosive eruption injecting material directly into the cloud layer.  

Scenario (1) is effectively the `dust from the surface' mechanism described above.  Albeit, by explicitly considering volcanism a source of juvenile OH-bearing material is introduced, which could help overcome the short surface lifetime of some OH-bearing phases. The second and third possibilities take advantage of the dynamics of volcanism to shorten the distance, and thereby potentially increase the flux, of mineral sources of OH to the clouds by reducing the time the material spends at high temperature near Venus's surface.  Volcanism may also help loft material above the sluggish surface winds to an altitude where winds more readily carry dust higher \citep[e.g.,][]{linkin1986vega,zasova2007structure,peralta2017venus}. 

Modelling work by \citet{glaze1999_jgr} and \citet{airey2015_pss} suggests that it is possible for volcanic eruption columns on Venus to reach the cloud base, however it requires particular circumstances that may or may not be frequently met: in particular, a high elevation vent and a magma containing several wt\% water.  Whether such water-rich magmas exist on Venus is unknown; taking Earth as an analogue, magmas with $> 3\,\mathrm{wt}\%$ water occur only where subduction is introducing surface water back into the mantle, a tectonic mode that cannot have prevailed on Venus for hundreds of millions of years, if ever.  Voluminous water-rich explosive volcanism is also problematic given the tight constraints on the below-cloud the \ce{H2O} mixing ratio of $<60\,\mathrm{ppm}$.  At best therefore, these constraints would imply a highly stochastic delivery of volcanic material directly into the clouds.

The best case for a volcanic contribution to mineral buffering of Venus's cloud chemistry is therefore by enhancing background dust levels in the below-cloud atmosphere.  \citet{fegley1989_nature} estimate that a volcanic flux of $\sim1\,\mathrm{km^3\,yr^{-1}}$ is required to sustain atmospheric \ce{SO_2} at the levels observed.  This translates to a mass flux of $\sim{3\times10^{12}\,\mathrm{kg\,yr^{-1}}}$ of magma, or $\sim6\times10^{11}\,\mathrm{mol\,yr^{-1}}$ of hydrated phases assuming terrestrial-levels of water in Venus's magmas.  As an upper bound, this volcanic flux could provide a potential $4\times10^{-15}\,\mathrm{mol\,cm^{-2}\,s^{-1}}$ of hydrated phases to Venus's clouds if the entire mass was mobilised as dust in the atmosphere.  Being below the $\sim{10^{-13}\,\mathrm{mol\,cm^{-2}\,s^{-1}}}$ of salt delivery estimated above, rates of volcanism either need to be (at least) an order of magnitude larger than assumed here, or the magmas correspondingly more volatile rich, for volcanism to be contributing to chemical buffering of Venus's clouds by water delivery.  We note that although seeming unlikely, given all the current uncertainties on the composition and dynamics of Venus's interior, this possibility cannot be entirely ruled out.

\subsection{Reconciling the Cloud Chemistry with Cloud Observations}
\label{sec:cloud-obs}

There is considerable evidence that the clouds of Venus are mostly sulfuric acid or something very like sulfuric acid. The classic paper by \citet{Young1973} identifies most of the lines of evidence. The refractive index of the clouds obtained from infrared polarimetry, constrained to within 1.425 and 1.455 at the time, is best explained by droplets of $\sim 75\%$ sulfuric acid. In addition, the bottom of the cloud layer matches the condensation temperature of sulfuric acid, and specific spectral features between 8 -- 13 $\mu$m are very similar to spectra of condensed sulfuric acid. Sulfuric acid is also expected at concentrations of $\sim 75\%$ based on models of \ce{H_2SO_4} and \ce{H_2O} condensation \citep{Kras2015}. Subsequent studies have further refined the estimated concentrations of sulfuric acid in the clouds. \citet{Barstow2012} perform a retrieval on VIRTIS data of the atmosphere from Venus Express, and find that the 2.2:1.74 $\mu$m radiance ratio is sensitive only to the imaginary index of refraction, and therefore the sulfuric acid concentrations, and that most of the retrieved sulfuric acid concentrations in the lower clouds are between 85 -- 96 wt\%, or between  16 -- 18 mol/L. \citet{Arney2014} used the same 2.2:1.74 $\mu$m radiance ratio and found that the sulfuric acid concentrations vary by time and latitude between 73 -- 87 wt\% in the upper clouds, or between 14 -- 16 mol/L sulfuric acid concentration, lower than \citet{Barstow2012}, suggesting that the concentration of sulfuric acid changes as a function of height. \citet{Titov2018} provide a comprehensive review of the research into Venus's clouds. None of the UV absorption due to our predicted profiles of gas-phase species, for the cloud model or any other model, explains the mysterious UV absorber, at $\lambda > 200$ nm. However, we do not consider the absorption properties of the aerosol particles themselves. It would be useful to determine the UV optical constants of these aerosols, or of sulfuric acid droplets themselves, to see whether a feasible candidate for the mysterious UV absorber is already in our midst.

Recent retrievals rely on infrared bands that constrain the index of refraction. If our proposed cloud chemistry is accurate and if all droplets have the same chemistry, hydrated sulfites and sulfates compose a large fraction of Venus's clouds. Sulfate achieves an index of refraction of $>1.44$ at $<20$ wt\% \ce{H_2O} \citep{Cotterell2017}, sufficient to explain the observed index of refraction. In addition, the \ce{S-H} and \ce{S-O} bonds are all similar, and so similar spectral features are expected. We predict a significant fraction of the clouds is made up of sulfites, in line with terrestrial \ce{SO_2} + \ce{H_2O} chemistry \citep{Terraglio1967}. The refractive index of sulfites is not as well known, though it has been measured for cyclic sulfites to be $\sim 1.5$ \citep{Pritchard1961}, which is consistent with Venus cloud observations, particularly those that favor values higher than can be achieved by pure sulfuric acid \citep{Markiewicz2018,Petrova2018}, but may be achievable by sulfites and sulfates.

One other possibility is that the droplets of Venus do not all have the same chemistry. The salt content of the aerosols could be heterogeneous, either because they derive from different surface, volcanic or delivered materials, or because some cloud nuclei are produced photochemically, as sulfur allotropes and sulfuric acid. The observed bimodal distribution of cloud droplet sizes \citep{Wilquet2009,Wilquet2012}, is consistent with this hypothesis. It may be that the smaller more numerous cloud droplets lack salts, while less numerous droplets have salts. The salts will afford those droplets a far greater capacity for \ce{SO_2}, and this may explain their larger size. If this is the case, we would predict the small mode droplets have a pH $< 0$ and the large mode droplets a pH $> 1$.

\subsection{The Effect of Cloud Chemistry on Gas-Phase Halogens}
\label{sec:cloud-hal}

Hydrogen halides have been observed on Venus, namely \ce{HCl} and \ce{HF} (see Appendix \ref{app:obs}). Attempts have been made to observe \ce{HBr}, but thus far only upper limits have been established \citep{Kras2016b}. These hydrogen halides are likely to dissolve into the droplets, especially if the pH is higher than previously thought, and this may affect both the droplet chemistry and the atmospheric profiles of these halides. We investigate this possibility, considering only \ce{HCl}, the sole hydrogen halide included in our model.

First, we have to determine the Henry's law constant for \ce{HCl} dissolved into water/sulfuric acid mixtures. \citet{Williams1993} find that the effective Henry's Law of \ce{HCl} in a solution of 50 wt\% \ce{H2SO4} is:
\begin{equation}
H^* \approx 10^5 \, {\rm M/bar},
\end{equation}
at room temperature (which we will use as a proxy for the cloud layer). We use the relation (where $H^*$ (mol/(L bar)) is the effective Henry's law constant and $K_a$ is the acid dissociation constant):
\begin{equation}
H^* = H \Big(1 + \dfrac{K_a}{a_{H^+}}\Big),
\end{equation}
to solve for the hydrogen activity, $a_{H^+}$. The concentration of 50 wt\% \ce{H2SO4} is approximately 2 mol/L, which will completely dissociate, and result in a pH of roughly -0.5, or an $a_{H^+} \approx 2$. For \ce{HCl}, the $K_a \approx 2 \times 10^6$. This means that $H \approx 0.1$ for \ce{HCl} in sulfuric acid.

We can now predict the depletion of \ce{HCl} for the sulfuric acid case, where pH $\sim -3$. The equation for the capacity, $\kappa$, of the cloud droplets for a particular species, assuming that the dissociation products are not removed and so a simple equilibrium is achieved, is:
\begin{align}
\kappa &= \dfrac{H^* p(\ce{HCl}) N_A}{n(\ce{HCl})} \, \int 4 \pi r_d^2 \, \Bigg(r_d \dfrac{dn_d}{dr_d}\Bigg) \, dr_d\\
&= H \Bigg(1 + \dfrac{K_a}{a_{H^+}}\Bigg) \, kTN_A \, \int \dfrac{4}{3} \pi r_d^2 \, \Bigg(r_d \dfrac{dn_d}{dr_d}\Bigg) \, dr_d. \label{eqn:capacity}
\end{align}
Though it is true that, if there is any \ce{Na^+} or other cation, it will form a salt with \ce{Cl^-}, this will act in solution as though entirely dissociated, and so effectively the amount of \ce{Cl-} and \ce{H^+} due to \ce{HCl} remains unchanged. If $\kappa < 1$, then the capacity of the droplets is not great enough to appreciably affect the composition of the gas-phase species. In the pure sulfuric acid case, $c \approx 1.6 \times 10^{-5}$, and so very little \ce{HCl} will be in the droplets, compared to the amount in the gas-phase.

At higher pH, the situation changes, but still, at pH $\sim 1$, $\kappa \approx 0.16$, and so there should not be significant depletion. The value of $\kappa$ in Equation (\ref{eqn:capacity}) can be set to one, and $a_{H^+}$ solved for to predict the pH at which HCl should be significantly depleted, pH $\approx 1.8$. Though the profiles of gas-phase hydrogen halides should not not be affected at lower pH's, the concentrations of halogens in the droplets could be significant, and it would be informative to measure these concentrations {\it in situ}.

\subsection{The Presence and Lifetime of Phosphine}
\label{sec:ph3}

A broad time-variable feature has been observed at 267 GHz by both ALMA and JCMT \citep{Greaves2020,Greaves2020b}. The existence and source of the feature has been disputed \citep{Snellen2020,Thompson2020}, though see the reply by \citet{Greaves2020b}. This feature has been attributed to phosphine \citep{Greaves2020,Greaves2020b}, consistent with possible {\it in situ} detection of phosphine discovered during re-analysis of Venus Pioneer data \citep{Mogul2021}. The updated ALMA data indicates that the 267 GHz feature is now consistent either with 1.5-7 ppb phosphine (\ce{PH_3}) or $\sim 50$ ppb mesospheric \ce{SO_2} \citep{Greaves2020b, Lincowski2021}, through this amount of \ce{SO_2} is not consistent with the non-detection of the 267.5 GHz \ce{SO_2} feature, from the same ALMA data, providing a 10 ppb upper limit \citep{Greaves2020b}. Careful modelling of different \ce{PH_3} profiles by \citet{Lincowski2021} demonstrate that the proposed in-cloud profile for \ce{PH_3} from \citet{Greaves2020} is insufficient to explain the observed feature. This leaves open the possibility that the signal is due to $\sim 50$ ppb mesospheric \ce{SO_2}, with an anomalous velocity shift, or mesospheric \ce{PH_3} in quantities difficult to reconcile with the \ce{PH_3} lifetime, unless there is an unknown mesospheric source of \ce{PH_3}. Further observations will be needed to determine which if either \ce{PH_3} or \ce{SO_2}, or if some other unidentified molecule is the source of the 267 GHz feature.

\citet{Greaves2020} and \citet{Bains2020} estimated the lifetime and required flux of 20 ppb \ce{PH_3}. The amount inferred from the ALMA observations has decreased to 1 ppb \citep{Greaves2020b}, and we can apply our new model results to estimate that a flux of $10^7$ cm$^{-2}$ s$^{-1}$ \ce{PH_3} is required to explain \ce{PH_3} at these abundances, assuming the 267 GHz feature probed the clouds (this also includes effective dry deposition of \ce{PH_2} set to fix the abundance of \ce{PH_2} at the surface equal to zero, and dry deposition of \ce{PH_3} fixed to $10^{-4} {\rm \, cm \, s^{-1}}$). We also find that \ce{PH_3} is efficiently destroyed above 60 km, and the steep gradient in its profile is consistent with the non-detection above 61 km \citep{Encrenaz2020}. The in-cloud \ce{PH_3} profile cannot explain the observed feature \citep{Lincowski2021}, but this profile is expected if there were a source of \ce{PH_3} in the clouds.

\subsection{Implications for Hypothetical Life in the Clouds of Venus}
\label{sec:implications-life}

The implications of this cloud chemistry on hypothetical Venusian life depends on how and in what form the sulfur is depleted. If surface, volcanic or exogenous delivery is responsible, this would explain the provision of various alkaline salts and/or other metals, essential for life as we know it. The higher pH is within the range where known acidophiles can thrive \citep{Messerli2005}. However, although acidophilic and halophilic extremophiles exist on Earth, there are, as far as we know, no known extremophiles that are both acidophilic and halophilic enough to thrive in these conditions \citep{Belilla2019}.

If on the other hand the pH is being buffered within the clouds by the organisms themselves, either by use of phosphine or by burning hydrocarbons, sacrificing themselves so that others may live, then the available water will be produced and scavenged by means of the reaction. The remaining question is whether the biomass is sufficient to explain the sulfur depletion, and there is no reliable estimate of the biomass in the clouds of Venus. Early estimates based on tentative detections of phosphine have been made \citep{Lingham2020}, but there is significant work left to better constrain the possibility that life is making use of the in-cloud \ce{SO_2}.

\section{Conclusion} 
\label{sec:conclusion}

In this paper, we discussed the puzzle of \ce{SO_2} depletion in the cloud layer. If the below-cloud observations of \ce{SO_2} and \ce{H_2O} are correct, then there is too little \ce{H_2O} to explain this depletion. We found that increasing the amount of below-cloud \ce{H_2O} predicts chemistry above the clouds that does not agree with observations, but decreasing the below-cloud \ce{SO_2} results in above-cloud chemistry that generally agrees with observations. We also explored the possibility that hydrogen is delivered into the clouds in the form of aerosols, salts or metals, either from an exogenous source, from dust rising up from the surface, from volcanism, and from processes occurring within the clouds themselves. These processes buffer the pH of the clouds to values of 1-2. We discuss the implication of these predictions for observations of other trace gas-phase species and the optical properties of the clouds themselves.

Probes into the clouds of Venus will be necessary to determine what is happening within the clouds: the depletion of \ce{SO_2}, the droplet chemistry (whether or not this chemistry has anything to do with the \ce{SO_2} depletion), the mysterious UV absorber, the known presence of heavy metals such as iron, the plausible presence of several reduced species in surprisingly large quantities \citep{Greaves2020,Mogul2021}. In particular, the DAVINCI+ mission concept is planned to travel into and below the clouds to measure atmospheric redox and better constrain the chemical cycles that are thought to sustain the clouds \citep{Garvin2020}. The cloud chemistry itself can be examined by including a design like the ``JPL Venus Aerosol Mass Spectrometer Concept'', where a nebulizer is incorporated onto a mass spectrometer to separate gas and cloud particles and analyze the aerosol chemistry directly \citep{Baines2018}. This would be the most straight-forward way to test this hypothesis, and other hypotheses, e.g. \citet{Kras2017}, that involve cloud chemistry.

Probes to the surface will be relevant for constraining the surface mineralogy and determining whether Venus's surface composition is in chemical equilibrium. These observations can be combined with a climate history of Venus, based on observations and models, to discover more about Venus's atmospheric evolution. Specifically, Figure~\ref{fig:GGchem} can also be used to speculate about the possible cause for the origin of the thick Venusian atmosphere that we observe today. If Venus once was a cooler planet with a thinner \ce{N2} dominated atmosphere, just like Earth, but for some reason it warmed up to temperatures above 580\,K, possibly due to large amounts of \ce{CO_2} released during global resurfacing \citep{Strom1994}, all carbonates in the surface rock would decompose and liberate even more \ce{CO2} into the atmosphere.  Not only would this increase the greenhouse effect, but it would also make the Venusian atmosphere thicker. Both effects would have increased the surface temperature, leading to a run-away build-up of the thick \ce{CO2} Venusian atmosphere that we find today. Observations of surface minerals, possible with VERITAS \citep{Smrekar2020}, and EnVision \citep{deOliveira2018}, would allow us to test the predictions of this and other hypotheses for the present state of Venus's atmosphere and climate, in particular, whether the surface and atmosphere of Venus are at thermochemical equilibrium.

To prepare for these missions, experiments are needed in Venus analogue environments to predict the chemistry that takes place, especially the largely unexplored chemistry that may take place within high concentrations of sulfuric acid, and the surface chemistry that may take place on efflorescent sulfates. In the meantime, a detailed cloud model of the form published by \citet{Gao2014}, but that incorporates this chemistry, would be of value in order to see if any remote predictions would distinguish between sulfuric acid/water droplets and sulfate/sulfite/water droplets. It may be possible to falsify the cloud chemistry hypothesis based on a combination of cloud formation and radiative transfer models and observations. It is unlikely the puzzles addressed in our paper are likely to be resolved without returning to the clouds of Venus. If other missions to other planets are any indication, what we find will be entirely unexpected.

\acknowledgments

The authors thank Joanna Petkowska-Hankel for the preparation of Fig. \ref{fig:cartoon}. P.~B.~R. thanks the Simons Foundation for funding (SCOL awards 599634). P.~W. acknowledges funding from the European Union H2020-MSCA-ITN-2019 under Grant Agreement no. 860470 (CHAMELEON). We thank David Grinspoon, Stephen Mojzsis and Kevin Zahnle for helpful discussions, Alex~T. Archibald for his advice for improving the Earth-relevant reactions for our network, and the entire team involved with \citet{Greaves2020} for initiating this investigation into the atmospheric chemistry of Venus. We also thank the two anonymous referees for helpful and insightful comments that helped improve the quality of this manuscript. A.~P. thanks Sami Mikhail for providing Venus and Venus II books and Christiane Helling for the support.

\vspace{5mm}

\software{\textsc{Argo} \citep{Rimmer2016}, \textsc{GGchem} \citep{Woitke2018}}

\appendix

\section{Observational Constraints on the Atmospheric Composition of Venus}
\label{app:obs}

The species we included in our network robustly, and that are also observationally constrained in Venus's atmosphere are carbon dioxide (\ce{CO_2}), molecular nitrogen (\ce{N_2}), sulfur dioxide (\ce{SO_2}), water vapor (\ce{H_2O}), carbon monoxide (\ce{CO}), molecular oxygen (\ce{O_2}), carbonyl sulfide (\ce{OCS}), sulfuric acid vapor (\ce{H_2SO_4}), hydrogen chloride (\ce{HCl}), sulfur monoxide (\ce{SO}), trisulfur (\ce{S_3}), tetrasulfur (\ce{S_4}), hydrogen sulfide (\ce{H_2S}) and molecular hydrogen (\ce{H_2}). Phosphine (\ce{PH_3}) is also included in our network but not in a robust way. \ce{PH_3} may have been observed in the atmosphere of Venus (see Section \ref{sec:ph3} for details). There are also many remote observations that constrain cloud properties such as average particle size and indirectly infer that the clouds are made of droplets of high concentration sulfuric acid (See Section \ref{sec:cloud-obs}). To this date no definitive {\it in situ} measurements of the cloud droplet chemistry have been made, and so virtually nothing is directly known about the cloud droplet chemistry. Observations have been made by a variety of instruments on the ground, by orbital probes, and by {\it in situ} probes. We compiled this data ourselves from a variety of sources, and a more complete compilation has been made by \citet{Johnson2019}, which includes several reactive species not incorporated into our network, such as \ce{HF}, as well as unreactive species. Our compilation is given in Table \ref{tab:obs}.

\startlongtable
\begin{deluxetable*}{lcccrr}
\tablecaption{Observational Constraints of Chemical Species in the Atmosphere of Venus\label{tab:obs}}
\tablehead{
\colhead{Species} & \colhead{$h_{\rm min}$} & 
\colhead{$h_{\rm max}$} & \colhead{Mixing Ratio$^*$} & 
\colhead{Reference} & \colhead{Obs Type} \\ 
\colhead{} & \colhead{(km)} & \colhead{(km)} & \colhead{} & 
\colhead{} & \colhead{}
} 
\startdata
{\bf \ce{SO_2}} & \nodata & \nodata & ppm & \nodata & \nodata \\
\nodata &       30.0    &       40.0    &       $130.   \pm     50.     $ &     \citet{Marcq2008}       &       Venus Express \\
\nodata &       0.      &       22.0    &       $130.   \pm     35.     $ &     \citet{Gelman1979}      &       Venera 12 \\
\nodata &       22      &       22      &       $185.   \pm     43.     $ &     \citet{Oyama1979}       &       Pioneer Venus \\
\nodata &       35      &       45      &       $130.   \pm     40.     $ &     \citet{Bezard2007}      &       Ground \\
\nodata &       12      &       12      &       $22.5   \pm     2.5     $ &     \citet{Bertaux1996}     &       Vega 1, Vega 2 \\
\nodata &       22      &       22      &       $38.0   \pm     3.8     $ &     \citet{Bertaux1996}     &       Vega 1, Vega 2 \\
\nodata &       42      &       42      &       $132.5  \pm     14.     $ &     \citet{Bertaux1996}     &       Vega 1, Vega 2 \\
\nodata &       52      &       52      &       $107.5  \pm     42.5    $ &     \citet{Bertaux1996}     &       Vega 1, Vega 2 \\
\nodata &       35      &       45      &       $130.   \pm     40.     $ &     \citet{Bezard2007}      &       Ground \\
\nodata &       42      &       42      &       $180.   \pm     70.     $ &     \citet{Pollack1993}     &       Ground \\
\nodata &       62      &       62      &       $0.5    \pm     0.1     $ &     \citet{Zasova1993}        &       Ground \\
\nodata &       62      &       62      &       $1.5    \pm     0.5     $ &     \citet{Zasova1993}        &       Ground \\
\nodata &       22      &       22      &       $38.    \pm     10.     $ &     \citet{Bertaux1996}     &       Vega 1 \\
\nodata &       12      &       12      &       $25.    \pm     10.     $ &     \citet{Bertaux1996}     &       Vega 1 \\
\nodata &       52      &       52      &       $150.   \pm     70.     $ &     \citet{Bertaux1996}     &       Vega 1 \\
\nodata &       52      &       52      &       $65.    \pm     30.     $ &     \citet{Bertaux1996}     &       Vega 2 \\
\nodata &       42      &       42      &       $125.   \pm     50.     $ &     \citet{Bertaux1996}     &       Vega 1 \\
\nodata &       42      &       42      &       $200.   \pm     100.    $ &     \citet{Bertaux1996}     &       Vega 2 \\
\nodata &       42      &       42      &       $< 176. $               &       \citet{Oyama1980}       &       Pioneer Venus \\
\nodata &       22      &       22      &       $185 \pm 43.1$               &      
\citet{Oyama1980}       &       Pioneer Venus \\
\nodata &       42      &       42      &        $324 \pm 148$              &       \citet{Oyama1980}       &       Pioneer Venus \\
\nodata &       52      &       52      &       $< 600. $               &       \citet{Oyama1980}       &       Pioneer Venus \\
\nodata &       22      &       22      &       $< 300. $               &       \citet{Hoffman1980}     &       Pioneer Venus \\
\nodata &       52      &       52      &       $< 10.  $               &       \citet{Hoffman1980}     &       Pioneer Venus \\
\nodata &       30      &       40      &       $130.   \pm     50.     $ &     \citet{Marcq2008}       &       Venus Express \\
\nodata &       88      &       88      &       $0.012  \pm     0.003   $ &     \citet{Encrenaz2012}    &       Ground \\
\nodata &       70      &       70      &       $0.075  \pm     0.025   $ &     \citet{Encrenaz2012}    &       Ground \\
\nodata &       70      &       70      &       $0.125  \pm     0.050   $ &     \citet{Na1990}          &       Ground \\
\nodata &       45      &       55      &       $140.   \pm     37.     $ &     \citet{Arney2014}       &       Ground \\
\nodata &       45      &       55      &       $126.   \pm     32.     $ &     \citet{Arney2014}       &       Ground \\
\nodata &       65      &       75      &       $0.38   \pm     0.07    $ &     \citet{Na1990}          &       IUE \\
\nodata &       65      &       75      &       $0.05   \pm     0.02    $ &     \citet{Na1990}          &       IUE \\
\nodata &       75      &       85      &       $50.0   \pm     10    $ &     \citet{Greaves2020}          &    Ground$^{**}$ \\
{\bf \ce{SO_2}} & \nodata & \nodata & ppm & \nodata & \nodata \\ \hline
{\bf \ce{H_2O}} & \nodata & \nodata & ppm & \nodata & \nodata \\
\nodata & 30    & 40    & $31   \pm 2           $ & \citet{Marcq2008}           & Venus Express \\
\nodata & 30    & 40    & $30   \pm 10          $ & \citet{deBergh1995}         & Venus Express \\
\nodata & 5     & 40    & $30   \pm 10          $ & \citet{Bezard2007}          & Venus Express \\
\nodata & 51.3  & 51.3  & $6.3  \pm 4.          $ & \citet{Donahue1997}         & Pioneer Venus \\
\nodata & 55.3  & 55.3  & $4.2  \pm 3.          $ & \citet{Donahue1997}         & Pioneer Venus \\
\nodata & 69.5  & 69.5  & $3.   \pm 1.          $ & \citet{Cottini2012}         & Venus Express \\
\nodata & 15    & 30    & $10.  \pm 1.          $ & \citet{Evans1969}           & Ground \\
\nodata & 49    & 58    & $700. \pm 300.        $ & \citet{Mukhin1982}          & Venera 13, 14 \\
\nodata & 46    & 50    & $2000.\pm 400.        $ & \citet{Surkov1982}          & Venera 13, 14 \\
\nodata & 26    & 45    & $30.  \pm 10.         $ & \citet{deBergh1995}         & Ground \\
\nodata & 15    & 30    & $30.  \pm 10.         $ & \citet{deBergh1995}         & Ground \\
\nodata & 0     & 15    & $30.  \pm 10.         $ & \citet{deBergh1995}         & Ground \\
\nodata & 10    & 10    & $45.  \pm 10.         $ & \citet{Meadows1996}         & Ground \\
\nodata & 0     & 10    & $7.   \pm 3.          $ & \citet{Donahue1992}         & Pioneer Venus \\
\nodata & 10    & 26    & $28.  \pm 18.         $ & \citet{Donahue1992}         & Pioneer Venus \\
\nodata & 58    & 60    & $20.  \pm 10.         $ & \citet{Moroz1990}           & Ground \\
\nodata & 52    & 60    & $1000.\pm 500.        $ & \citet{Surkov1987}          & Vega 1, Vega 2 \\
\nodata & 42    & 42    & $150. \pm 50.         $ & \citet{Moroz1979}           & Venera 11, 12 \\
\nodata & 22    & 22    & $60.  \pm 30.         $ & \citet{Moroz1979}           & Venera 11, 12 \\
\nodata & 35    & 45    & $40.  \pm 20.         $ & \citet{Bezard1990}          & Ground \\
\nodata & 10    & 40    & $30.  \pm 10.         $ & \citet{Pollack1993}         & Ground \\
\nodata & 15    & 25    & $30.  \pm 10.         $ & \citet{deBergh1995}         & Ground \\
\nodata & 0     & 15    & $30.  \pm 15.         $ & \citet{deBergh1995}         & Ground \\
\nodata & 0     & 0     & $20.  \pm 10.         $ & \citet{Moroz1979}           & Venera 11, 12 \\
\nodata & 30    & 40    & $26.  \pm 4.          $ & \citet{Marcq2006}           & Ground \\
\nodata & 30    & 40    & $35.  \pm 4.          $ & \citet{Tsang2008}           & Venus Express \\
\nodata & 30    & 40    & $30.  \pm 4.          $ & \citet{Tsang2008}           & Venus Express \\
\nodata & 30    & 40    & $34.  \pm 2.          $ & \citet{Arney2014}           & Ground \\
\nodata & 30    & 40    & $33.  \pm 3.          $ & \citet{Arney2014}           & Ground \\
\nodata & 15    & 30    & $33.  \pm 2.          $ & \citet{Arney2014}           & Ground \\
\nodata & 15    & 30    & $32.  \pm 2.          $ & \citet{Arney2014}           & Ground \\
\nodata & 0     & 15    & $44.  \pm 9.          $ & \citet{Bezard2007}          & Venus Express \\
\nodata & 0     & 15    & $30.  \pm 10.         $ & \citet{Bezard2011}          & Venus Express \\
\nodata & 0     & 15    & $31.  \pm 9.          $ & \citet{Chamber2013}         & AAT \\
\nodata & 0     & 15    & $29.  \pm 2.          $ & \citet{Arney2014}           & Ground \\
\nodata & 0     & 15    & $27.  \pm 2.          $ & \citet{Arney2014}           & Ground \\
\nodata & 0     & 15    & $25.7 \pm 1.4         $ & \citet{Fedorova2015}        & Venus Express \\
\nodata & 0     & 15    & $29.4 \pm 1.6         $ & \citet{Fedorova2015}        & Venus Express \\
\nodata & 65    & 74    & $6.0  \pm 4.0         $ & \citet{Fedorova2016}        & Pioneer Venus \\
\nodata & 70    & 110   & $1.0  \pm 0.9         $ & \citet{Fedorova2008}        & Venus Express \\
\nodata & 45    & 55    & $20.0 \pm 10.         $ & \citet{Meadows1996}         & Ground \\
\nodata & 0     & 15    & $45.  \pm 15.         $ & \citet{Meadows1996}         & Ground \\
\nodata & 65    & 100   & $1.8  \pm 1.8         $ & \citet{Sandor2005}          & Ground \\
\nodata & 65    & 120   & $2.5  \pm 0.6         $ & \citet{Encrenaz2015}        & Ground \\
\nodata & 65    & 75    & $3.0  \pm 1.0         $ & \citet{Cottini2012}         & Venus Express \\
\nodata & 65    & 75    & $5.0  \pm 2.0         $ & \citet{Cottini2012}         & Venus Express \\
\nodata & 65    & 65    & $0.8         $ & \citet{Bell1991}         & Ground \\
\nodata & 55    & 55    & $100.0       $ & \citet{Bell1991}         & Ground \\
\nodata & 40    & 40    & $200.0       $ & \citet{Bell1991}         & Ground \\
{\bf \ce{H_2O}} & \nodata & \nodata & ppm & \nodata & \nodata \\ \hline
{\bf \ce{CO}} & \nodata & \nodata & ppm & \nodata & \nodata \\ 
\nodata & 36    & 36    & $27.5         \pm 3.5         $ & \citet{Marcq2008}   & Venus Express \\
\nodata & 35    & 35    & $23.0         \pm 2.0         $ & \citet{Tsang2008}   & Venus Express \\
\nodata & 35    & 35    & $32.0         \pm 2.0         $ & \citet{Tsang2008}   & Venus Express \\
\nodata & 22    & 22    & $20.0         \pm 0.4         $ & \citet{Oyama1979}   & Pioneer Venus \\
\nodata & 64    & 64    & $45.0         \pm 10.0        $ & \citet{Fegley2014}  & Venera 13, 14 \\
\nodata & 0     & 0     & $3.8          \pm 3.2         $ & \citet{Fegley2014}  & Venera 13, 14 \\
\nodata & 52    & 52    & $32.0         \pm 22.0        $ & \citet{Oyama1980}   & Pioneer Venus \\
\nodata & 42    & 42    & $30.0         \pm 18.0        $ & \citet{Oyama1980}   & Pioneer Venus \\
\nodata & 22    & 22    & $20.0         \pm 3.0         $ & \citet{Oyama1980}   & Pioneer Venus \\
\nodata & 0     & 42    & $28.0         \pm 7.0         $ & \citet{Gelman1979}  & Venera 12 \\
\nodata & 64.   & 64.   & $45.0         \pm 10.0        $ & \citet{Connes1968}  & Ground \\
\nodata & 64.   & 64.   & $51.0         \pm 1.0        $ & \citet{Young1972}   & Ground \\
\nodata & 90    & 90    & $180.         \pm 90.0        $ & \citet{Wilson1981}  & Ground \\
\nodata & 36    & 36    & $23.0         \pm 5.0         $ & \citet{Pollack1993} & Ground \\
\nodata & 40    & 40    & $29.0         \pm 7.0         $ & \citet{Pollack1993} & Ground \\
\nodata & 35    & 45    & $45.0         \pm 10.0        $ & \citet{Bezard1990}  & Ground \\
\nodata & 36    & 36    & $23.0         \pm 10.0        $ & \citet{Pollack1993} & Ground \\
\nodata & 28    & 28    & $23.0         \pm 10.0        $ & \citet{Bezard2007}  & Ground \\
\nodata & 42    & 42    & $30.0         \pm 15.0        $ & \citet{Bezard2007}  & Ground \\ 
\nodata & 36    & 36    & $24.0         \pm 2.0         $ & \citet{Marcq2006}   & Ground \\
\nodata & 36    & 36    & $27.0         \pm 3.0         $ & \citet{Cotton2012}  & Ground \\
\nodata & 68    & 71    & $70.0         \pm 8.0        $ & \citet{Kras2008}    & Ground \\
\nodata & 68    & 68    & $51.0         \pm 4.0         $ & \citet{Kras2010}    & Ground \\
\nodata & 68    & 68    & $40.0         \pm 4.0         $ & \citet{Kras2010}    & Ground \\
\nodata & 104   & 104   & $560.0        \pm 100.0       $ & \citet{Kras2014}    & Ground \\
\nodata & 70    & 70    & $35.0         \pm 10.0        $ & \citet{Marcq2015}   & Ground \\
\nodata & 30    & 30    & $30.0         \pm 5.0         $ & \citet{Collard1993} & Galileo \\
\nodata & 30    & 30    & $40.0         \pm 5.0         $ & \citet{Collard1993} & Galileo \\
\nodata & 35    & 35    & $30.0         \pm 15.0        $ & \citet{Marcq2005}   & Ground \\
\nodata & 104   & 104   & $560.0        \pm 100.0       $ & \citet{Kras2014}    & Ground \\
\nodata & 65    & 75    & $70.0         \pm 10.0        $ & \citet{Grassi2014}  & Venus Express \\
\nodata & 65    & 75    & $60.0         \pm 5.0         $ & \citet{Grassi2014}  & Venus Express \\
{\bf \ce{CO}} & \nodata & \nodata & ppm & \nodata & \nodata \\ \hline
{\bf \ce{O_2}} & \nodata & \nodata & ppm & \nodata & \nodata \\
\nodata & 52    & 52    & $43.  \pm 25. $ & \citet{Oyama1979}   & Pioneer Venus \\
\nodata & 42    & 42    & $16.  \pm 8.  $ & \citet{Oyama1979}   & Pioneer Venus \\
\nodata & 52    & 52    & $44.  \pm 25. $ & \citet{Oyama1980}   & Pioneer Venus \\
\nodata & 42    & 42    & $16.  \pm 7.  $ & \citet{Oyama1980}   & Pioneer Venus \\
\nodata & 35    & 58    & $18.  \pm 4.  $ & \citet{Mukhin1982}  & Venera 13, 14 \\
\nodata & 60    & 100    & $<2.8$ & \citet{Marcq2018}  & Ground \\
{\bf \ce{O_2}} & \nodata & \nodata & ppm & \nodata & \nodata \\ \hline
{\bf \ce{OCS}} & \nodata & \nodata & ppm & \nodata & \nodata \\
\nodata & 30    & 30    & $14.0         \pm 6.0         $ & \citet{Pollack1993} & Ground \\
\nodata & 33    & 33    & $3.25         \pm 0.75        $ & \citet{Marcq2008}   & Venus Express \\
\nodata & 29    & 37    & $40.0         \pm 20.0        $ & \citet{Mukhin1982}  & Venera 13, 14 \\
\nodata & 33    & 33    & $4.4          \pm 1.0         $ & \citet{Pollack1993} & Ground \\
\nodata & 28    & 28    & $30.0         \pm 10.0        $ & \citet{Pollack1993} & Ground \\
\nodata & 38    & 38    & $0.35         \pm 0.1         $ & \citet{Marcq2005}   & Ground \\
\nodata & 0     & 30    & $15.0         \pm 5.0         $ & \citet{Bezard2007}  & Ground \\
\nodata & 30    & 30    & $14.0         \pm 6.0         $ & \citet{Pollack1993} & Ground \\
\nodata & 29    & 37    & $40.0         \pm 20.0        $ & \citet{Mukhin1983}  & Venera 13, 14 \\
\nodata & 36    & 36    & $0.52         \pm 0.05        $ & \citet{Marcq2006}   & Ground \\
\nodata & 30    & 30    & $16.0         \pm 8.0         $ & \citet{Bezard2007}  & Ground \\
\nodata & 38    & 38    & $0.35         \pm 0.1         $ & \citet{Bezard2007}  & Ground \\
\nodata & 36    & 36    & $0.44         \pm 0.1         $ & \citet{Arney2014}   & Ground \\
\nodata & 36    & 36    & $0.57         \pm 0.12        $ & \citet{Arney2014}   & Ground \\
\nodata & 36    & 36    & $0.5          \pm 0.02        $ & \citet{Marcq2005}   & Ground \\
\nodata & 36    & 36    & $0.46         \pm 0.01        $ & \citet{Marcq2005}   & Ground \\
\nodata & 36    & 36    & $0.54         \pm 0.13        $ & \citet{Arney2014}   & Ground \\
\nodata & 36    & 36    & $0.61         \pm 0.12        $ & \citet{Arney2014}   & Ground \\
\nodata & 65    & 65    & $<0.004                       $ & \citet{Kras2010}    & Ground \\
\nodata & 65    & 65    & $0.005        \pm 0.003       $ & \citet{Kras2010}    & Ground \\
{\bf \ce{OCS}} & \nodata & \nodata & ppm & \nodata & \nodata \\ \hline
{\bf \ce{H_2SO_4}} & \nodata & \nodata & ppm & \nodata & \nodata \\
\nodata & 50      & 52      & $3.0 \pm 2.0$    & \citet{Oschlisniok2012}        & Venus Express \\
{\bf \ce{H_2SO_4}} & \nodata & \nodata & ppm & \nodata & \nodata \\ \hline
{\bf \ce{HCl}} & \nodata & \nodata & ppm & \nodata & \nodata \\ 
\nodata & 15    & 25    & $0.41         \pm 0.04        $ & \citet{Arney2014}   & Ground \\
\nodata & 15    & 25    & $0.42         \pm 0.055       $ & \citet{Arney2014}   & Ground \\
\nodata & 74    & 74    & $0.4          \pm 0.03        $ & \citet{Kras2010b}   & Ground \\
\nodata & 61    & 67    & $0.76         \pm 0.1         $ & \citet{Iwagami2008} & Ground \\
\nodata & 74    & 74    & $0.4          \pm 0.03        $ & \citet{Kras2010b}   & Ground \\
\nodata & 70    & 70    & $0.4          \pm 0.04        $ & \citet{Sandor2012}   & Ground \\
\nodata & 90    & 90    & $0.2          \pm 0.02        $ & \citet{Sandor2012}   & Ground \\
\nodata & 35    & 70    & $0.5          \pm 0.12        $ & \citet{Connes1967}   & Ground \\
\nodata & 74    & 74    & $0.4          \pm 0.04        $ & \citet{Kras2010b}   & Ground \\
\nodata & 36    & 36    & $0.42         \pm 0.07        $ & \citet{Young1972}   & Ground \\
\nodata & 50    & 50    & $0.61         \pm 0.06        $ & \citet{Young1972}   & Ground \\
\nodata & 30    & 40    & $0.005        \pm 0.002       $ & \citet{Bezard1990}  & Ground \\
\nodata & 70    & 70    & $0.4          \pm 0.04        $ & \citet{Bezard1990}  & Ground \\
\nodata & 15    & 30    & $0.5          \pm 0.05        $ & \citet{Bezard1990}  & Ground \\
\nodata & 45    & 55    & $0.5          \pm 0.05        $ & \citet{Iwagami2008} & Ground \\
\nodata & 65    & 75    & $0.6          \pm 0.1         $ & \citet{Connes1967}  & Ground \\
{\bf \ce{HCl}} & \nodata & \nodata & ppm & \nodata & \nodata \\ \hline
{\bf \ce{PH_3}} & \nodata & \nodata & ppb & \nodata & \nodata \\
\nodata & 55.0  & 60.0  & $20.0 \pm 10.0$  & \citet{Greaves2020} & Ground$^{\dagger}$ \\
\nodata & 55.0  & 60.0  & $1.5 \pm 1.0$  & \citet{Greaves2020b} & Ground$^{\dagger}$ \\
\nodata & 60.0  & 65.0  & $< 5.0$  & \citet{Encrenaz2020} &
Ground \\
{\bf \ce{PH_3}} & \nodata & \nodata & ppb & \nodata & \nodata \\ \hline
{\bf \ce{SO}} & \nodata & \nodata & ppb & \nodata & \nodata \\
\nodata & 84.0  & 90.0  & $ 8.0 \pm 2.0         $ & \citet{Encrenaz2015}        & Ground \\
\nodata & 84.0  & 90.0  & $ 8.0 \pm 2.0         $ & \citet{Encrenaz2015}        & Ground \\
\nodata & 75.0  & 85.0  & $ 5.0 \pm 4.0        $ & \citet{Na1990}              & Ground \\
{\bf \ce{SO}} & \nodata & \nodata & ppb & \nodata & \nodata \\ \hline
{\bf \ce{S_3}} & \nodata & \nodata & ppb & \nodata & \nodata \\
\nodata & 23.0  & 23.0  & $0.04         \pm 0.01        $ & \citet{Bezard2007}  & Venera 11-14 \\
\nodata & 3.0   & 19.0  & $0.065        \pm 0.035       $ & \citet{Maiorov2005} & Venera 11 \\
\nodata & 3.0   & 19.0  & $11.          \pm 3.0         $ & \citet{Kras2013}    & Venera 11 \\
\nodata & 10.0  & 19.0  & $18.          \pm 3.0         $ & \citet{Kras2013}    & Venera 11 \\
{\bf \ce{S_3}} & \nodata & \nodata & ppb & \nodata & \nodata \\ \hline
{\bf \ce{S_4}} & \nodata & \nodata & ppb & \nodata & \nodata \\
\nodata & 3.0   & 10.0  & $4.   \pm 3.9 $ & \citet{Kras2013}    & Venera 11 \\
\nodata & 10.0  & 19.0  & $6.   \pm 2.0 $ & \citet{Kras2013}    & Venera 11 \\
{\bf \ce{S_4}} & \nodata & \nodata & ppb & \nodata & \nodata \\ \hline
{\bf \ce{H_2S}} & \nodata & \nodata & ppb & \nodata & \nodata \\
\nodata & 70.0  & 70.0  & $<13.0$  & \citet{Kras2008} & Ground \\
{\bf \ce{H_2S}} & \nodata & \nodata & ppb & \nodata & \nodata \\ \hline
{\bf \ce{H_2}} & \nodata & \nodata & ppm & \nodata & \nodata \\
\nodata & 0.0  & 25.0  & $<10.$  & \citet{Donahue1997} & Pioneer Venus \\
\nodata & 22.0  & 64.0  & $<10.$  & \citet{Oyama1980} & Pioneer Venus \\
\nodata & 49.0  & 58.0  & $25. \pm 10$  & \citet{Mukhin1982} & Venera 13,14 \\
{\bf \ce{H_2}} & \nodata & \nodata & ppm & \nodata & \nodata
\enddata
\vspace{1mm}
$^*$ Values and errors as found in the cited literature.\\
$^{**}$ If interpreted as mesospheric \ce{SO_2}, see Section \ref{sec:ph3}\\
$^{\dagger}$ If interpreted as mesospheric \ce{PH_3}, see Section \ref{sec:ph3}\\
$^{\ddagger}$ Based on their wet profile, which fits their data best.
\end{deluxetable*}

\section{The Chemical Network: Stand2020}
\label{app:stand}

We constructed our network starting with \citet{Rimmer2019}, adjusting several reactions relevant for Earth's atmosphere to improve agreement between our model predictions and observations of formaldehyde and \ce{HCN} in modern Earth's atmosphere. We then include the sulfur network developed by \citet{Hobbs2020}, and supplement it with reactions from \citet{Kras2007} and \citet{Zhang2012}. We made extensive use of the NIST database \citep{Manion2015}, the KIDA database \citep{Wakelam2012}, the MPI-Mainz UV/VIS database \citep{Keller2013}, and PhiDRates \citep{Huebner2015}, keeping with the philosophy, quoted from \citet{Rimmer2016}:

\begin{enumerate}
\item If there exists only one published rate constant for a given reaction, we use that value.
\item Reject all rate constants that become unrealistically large at extreme temperature.
\item Choose rate constants that agree with each other over the range of validity.
\item If the most recent published rate constant disagrees with (3), and the authors give convincing arguments
for why the previous rates were mistaken, we use the most recently published rate.
\end{enumerate}

The network is given in Tables \ref{tab:network}, \ref{tab:photochem} and \ref{tab:cond}.

\startlongtable
\begin{deluxetable*}{lccccr}
\tablecaption{Reactions Added and Updated for STAND2020\label{tab:network}}
\tablehead{
\colhead{Reaction} & \colhead{$\alpha$} & 
\colhead{Units} & \colhead{$\beta$} & 
\colhead{$E_a$ (K)} & \colhead{Reference}
} 
\startdata
$\ce{C} + \ce{C} + \ce{M} \rightarrow \ce{C_2} + \ce{M}$ & $5.46 \times 10^{-31}$ & cm$^{6}$ s$^{-1}$ & -1.6 & 0.0 & \citet{Slack1976}\\
$\ce{C} + \ce{C} + \ce{M} \rightarrow \ce{C_2} + \ce{M}$ & $2.16 \times 10^{-11}$ & cm$^{3}$ s$^{-1}$ & 0.0 & 0.0 & \citet{Martinotti1968}\\
$\ce{C} + \ce{O} + \ce{M} \rightarrow \ce{CO} + \ce{M}$ & $2.00 \times 10^{-34}$ & cm$^{6}$ s$^{-1}$ & 0.0 & 0.0 & \citet{Fairbairn1969}\\
$\ce{C} + \ce{O} + \ce{M} \rightarrow \ce{CO} + \ce{M}$ & $4.82 \times 10^{-15}$ & cm$^{3}$ s$^{-1}$ & -1.0 & 0.0 & Est.\\
$\ce{H} + \ce{N} + \ce{M} \rightarrow \ce{HN} + \ce{M}$ & $5.02 \times 10^{-32}$ & cm$^{6}$ s$^{-1}$ & 0.0 & 0.0 & \citet{Brown1973}\\
$\ce{H} + \ce{N} + \ce{M} \rightarrow \ce{HN} + \ce{M}$ & $1.21 \times 10^{-12}$ & cm$^{3}$ s$^{-1}$ & -1.0 & 0.0 & Est.\\
$\ce{N} + \ce{N} + \ce{M} \rightarrow \ce{N_2} + \ce{M}$ & $1.25 \times 10^{-32}$ & cm$^{6}$ s$^{-1}$ & 0.0 & 0.0 & \citet{Knipovich1988}\\
$\ce{N} + \ce{N} + \ce{M} \rightarrow \ce{N_2} + \ce{M}$ & $5.00 \times 10^{-16}$ & cm$^{3}$ s$^{-1}$ & 0.0 & 0.0 & \citet{Takahashi1977}\\
\enddata
\vspace{2mm}
$^a$ This reaction is not reversed.\\
$^b$ The rate constant for this reaction was miswritten in \citet{Kras2007,Zhang2012}.
\end{deluxetable*}

\startlongtable
\begin{deluxetable*}{lcr}
\tablecaption{Photochemistry Reactions in   STAND2020\label{tab:photochem}}
\tablehead{
\colhead{Reactants} & \colhead{Database} & \colhead{References}
} 
\startdata
\ce{C} & \textsc{PhiDRates} & \citet{Verner1995,Verner1996} \\
\hline
\ce{C(^1D)} & \textsc{PhiDRates} & \citet{Barfield1972,Henry1970} \\
\hline
\ce{C(^1S)} & \textsc{PhiDRates} & \citet{Barfield1972,Henry1970} \\
\hline
\ce{H} & \textsc{PhiDRates} & \citet{Stobbe1930,Sauter1931,Bethe1957} \\
\hline
\ce{He} & \textsc{PhiDRates} & \citet{Verner1995,Verner1996} \\
\hline
\ce{N} & \textsc{PhiDRates} & \citet{Verner1995,Verner1996} \\
\hline
\ce{O} & \textsc{PhiDRates} & \citet{Verner1995,Verner1996} \\
\hline
\ce{O(^1D)} & \textsc{PhiDRates} & \citet{Barfield1972,Henry1970} \\
\enddata
\vspace{2mm}
$^a$ \ce{HO} has both theoretical and experimental estimates, and one can switch between them. Experimental values are used for all results presented here.
\end{deluxetable*}

\startlongtable
\begin{deluxetable*}{lcr}
\tablecaption{Modified Antoine Equations for Vapor Pressure of Condensible Species\label{tab:cond}}
\tablehead{
\colhead{Species} & \colhead{Equation} & \colhead{References}
}
\vspace{2mm}
\startdata
\ce{CH_4} & $\log_{10} p = 4.4251 - \dfrac{453.92}{T} - \dfrac{4055.6}{T^2}$ & \vspace{2mm}\\
 & $+\dfrac{1.1535 \times 10^5}{T^3} + \dfrac{1.1656\times10^{6}}{T^4}$ & \citet{Moses1992} \vspace{2mm}\\
\ce{H_2O} & $\log_{10} p = 6.8543 - \dfrac{1897.0}{T - 64.848}$ & \citet{Rimmer2020} \vspace{2mm}\\
\ce{H_3N} & $\log_{10} p = 15.96 - \dfrac{3537.0}{T} - \dfrac{33100.0}{T^2}$ & \vspace{2mm} \\
 &  $+\dfrac{1.752 \times 10^6}{T^3} + \dfrac{2.995\times10^{7}}{T^4}$ & Est. \vspace{2mm}\\
\ce{HCN} & $\log_{10} p = 20.055 - \dfrac{4522.2}{T}$ & \citet{Sinozaki1926} \vspace{2mm}\\
\ce{N_2O} & $\log_{10} p = 16.22 - \dfrac{2971.0}{T}$ & \citet{Stull1947} \vspace{2mm}\\
\ce{C_2H_2} & $\log_{10} p = 6.2224 - \dfrac{1644.1}{T}$ & \citet{Ambrose1956,Ambrose1964} \vspace{2mm}\\
\enddata
\vspace{2mm}
$T$ is in units of kelvin, $p$ in units of bar.\\
$^a$ The isomer will matter significantly for vapor pressures. This is a way to arbitrarily constraint the abundances of large hydrocarbons in cold reducing atmospheres.\\
$^b$ Estimated to be the same as \ce{H_2SO_4} in the atmosphere of Venus.
\end{deluxetable*}

\bibliographystyle{aasjournal}

\end{document}